\documentclass[journal]{IEEEtran}
\usepackage{multirow}
\usepackage{cite}
\ifCLASSINFOpdf

\else

\fi
\usepackage{threeparttable}
\usepackage{amsmath}
\usepackage{url}
\usepackage{graphicx}
\usepackage{xcolor} 
\usepackage{amsmath} 
\usepackage{booktabs}
\usepackage{tablefootnote}
\usepackage{steinmetz}
\usepackage{soul}
\usepackage{subfigure}
\begin{document}

\title{PoLyScriber: Integrated Fine-tuning of Extractor and Lyrics Transcriber for Polyphonic Music}

\author{Xiaoxue Gao,~\IEEEmembership{Member,~IEEE,}
         Chitralekha Gupta,~\IEEEmembership{Member,~IEEE,}
        Haizhou Li,~\IEEEmembership{Fellow,~IEEE}
\thanks{Manuscript received October 1, 2022; revised March 1, 2023; accepted May 2, 2023. This work is supported by the Agency for Science, Technology and Research (A*STAR) under its AME Programmatic Funding Scheme (Project No. A18A2b0046), and IAF, A*STAR, SOITEC, NXP and National University of Singapore under FD-fAbrICS: Joint Lab for FD-SOI Always-on Intelligent \& Connected Systems (Award I2001E0053).}
\thanks{Xiaoxue Gao is with the Department of Electrical and Computer Engineering, National University of Singapore, Singapore 117583 (e-mail: xiaoxue@nus.edu.sg).}
\thanks{Chitralekha Gupta is with the School of Computing, National University of Singapore, Singapore 117417 (e-mail: chitralekha@nus.edu.sg).}
\thanks{Haizhou Li is with Shenzhen Research Institute of Big Data, School of Data Science, The Chinese University of Hong Kong, Shenzhen, China 518172, the Department of Electrical and Computer Engineering, National University of Singapore, Singapore 117583 (e-mail: haizhouli@cuhk.edu.cn).}
}

\maketitle

\begin{abstract}
Lyrics transcription of polyphonic music is challenging as the background music affects lyrics intelligibility. Typically, lyrics transcription can be performed by a two-step pipeline, i.e. a singing vocal extraction front end, followed by a lyrics transcriber back end, where the front end and back end are trained separately. Such a two-step pipeline suffers from both imperfect vocal extraction and mismatch between front end and back end. In this work, we propose a novel end-to-end integrated fine-tuning framework, that we call PoLyScriber, to globally optimize the vocal extractor front end and lyrics transcriber back end for lyrics transcription in polyphonic music. The experimental results show that our proposed PoLyScriber achieves substantial improvements over the existing approaches on publicly available test datasets.
\end{abstract}


\IEEEpeerreviewmaketitle

\section{Introduction}

\IEEEPARstart{L}{yrics} constitute the fundamental textual component of singing voice in music for the emotional perception of the song. Lyrics also helps in foreign language learning~\cite{good2015efficacy}. Music analysis has attracted great attention recently including audio-visual music source separation, localization \cite{zhao2018sound,zhao2019sound,gan2020music} and music generation \cite{gan2020foley,chen2022sound2synth}. Being able to understand the sung text may well contribute to listeners' enjoyment of music~\cite{collister2008comparison}. 
Lyrics transcription seeks to recognize the sung lyrics from singing vocals in the presence of instrumental music accompaniment. Various applications, such as automatic generation of karaoke lyrical content, music video subtitling~\cite{dzhambazov2017knowledge} and query-by-singing~\cite{hosoya2005lyrics} can benefit from automatic lyrics transcription.  

In lyrics transcription, we encounter many technical challenges. A prior study~\cite{fine2014making} shows that lyrics transcription in polyphonic music is challenging even for professional musicians, as the intelligibility is affected by many factors including complex structure of singing, diverse polyphonic music and different environmental conditions. Specifically, singing shows a higher degree of pronunciation and prosody variation than speech~\cite{fujihara2012lyrics,gao2020personalized,gao2019speaker,vijayan2018analysis}, therefore lyrics transcription is more challenging than automatic speech recognition (ASR). Moreover, background music accompaniment interferes with sung voice. The accompaniment typically adds an intricate and structured source of musical information to the singing vocals, and often influences the lyrics intelligibility of the singing vocal \cite{sharma2020automatic,ibrahim2017intelligibility}. Thus, any detrimental effects of singing to lyrics intelligibility are likely to be exacerbated by background music distraction.

To address the problem of automatic lyrics transcription in polyphonic music, there are mainly two broad approaches, 1) \textit{direct modeling} (DM) approach which takes the polyphonic audio, i.e. singing vocals + background music as model input directly to transcribe lyrics~\cite{stoller2019,9054567}, and 2) \textit{extraction-transcription} approach which extracts singing vocals first through a trained source separation model, and then transcribes the lyrics of the extracted vocals~\cite{gupta2019,mesaros2010automatic,dzhambazov2015modeling,fujihara2011lyricsynchronizer}.

In direct modeling approaches, acoustic models are trained directly on the polyphonic music, a mixture of vocals and music accompaniments. For example, Stoller et al.~\cite{stoller2019} developed a system based on a Wave-U-Net architecture to predict character probabilities directly from raw audio. This system performed well for the task of lyrics-to-audio alignment, however, showed a high word error rate (WER) in lyrics transcription. 
Music-informed acoustic modelling that incorporated music genre-specific information are proposed~\cite{9054567,gao2022genre}.
The study in \cite{9054567} suggested that lyrics acoustic models can benefit from music genre knowledge of the background music but it requires an additional genre extraction step, separately. Moreover, background music is difficult to model explicitly due to complicated rhythmic and harmonic structure, diverse music genres and compositional styles~\cite{fine2014making}. The effect of interference caused by the background music has not been well-studied under direct modeling approaches.

Humans are able to recognize lyrical words of a song by paying more attention to the singing vocal than the background music. The extraction-transcription approach~\cite{mesaros2010automatic,demirel2021low, demirel2020automatic} attempts to emulate such as human listening process through a two-step pipeline in which, first, the singing vocal is extracted from the polyphonic input signal, then the lyrics are transcribed from the extracted singing vocal. The two-step pipeline suffers from both imperfect vocal extraction and mismatch between the vocal extraction front end and the lyrics decoding back end~\cite{gupta2019,9054567}. In the two-step pipeline, the front end is not optimized for lyrics transcription, and the mismatch is partly because front end and back end are trained separately with different objective functions. 

In this paper, we propose an end-to-end (E2E) integrated fine-tuning solution, PoLyScriber, that integrate vocal extraction with lyrics transcription in a single network.  This is a departure from the previous studies~\cite{gupta2019,9054567,mesaros2010automatic,demirel2021low, demirel2020automatic}, where extraction and transcription are trained separately with mismatch problem. To the best of our knowledge, this work presents the first study on an integrated extraction-transcription solution to automatic lyric transcription, where the integrated fine-tuning is designed to fine-tune an integrated framework that consists of a pre-trained extractor and a pre-trained lyrics transcriber.
The contributions of this paper include: 1) novel E2E neural solutions for automatic lyrics transcription that combines the two-step pipeline into one through an integrated fine-tuning framework; 2) a systematic comparative study over several solutions and provide our insightful findings that the appropriate amount of removal of background accompaniments can be achieved by the integrated fine-tuning, where its intermediate vocal outputs lie in the middle of complete music removal and complete music presence.

The rest of this paper is organized as follows. Section \ref{Related Work} presents the related work to set the stage for this study. In Section~\ref{PoLyScriber}, we present the overview of our proposed PoLyScriber. Section~\ref{PoLyScriber Components} introduces the components of the proposed PoLyScriber in detail. Section~\ref{Experiments} consists of the database and experimental setup. Section~\ref{Results and Discussion} discusses the experiment results. Finally, Section~\ref{Conclusion} concludes the study.

\section{Related Work}
\label{Related Work}
We discuss the related work in singing voice separation as well as lyrics transcription of solo-singing and polyphonic music to motivate this study.

\subsection{Singing Voice Separation}
The extraction-transcription approach makes use of a singing voice separation (SVS) front end to separate singing vocals from polyphonic music.
There are generally two approaches for SVS, namely waveform-based and spectrogram-based approaches~\cite{gupta2022overview}. The waveform-based approach directly takes time-domain polyphonic music signal as input and separately output singing voice and music signals~\cite{stoller2018wave,lluis2018end,samuel2020meta,defossez2019music}. As the waveform-based encoder learns directly from input data, it is sensitive to the change of music content. The spectrogram-based approach applies frequency analysis in the front end.  
It predicts a power spectrum for each source and re-use the phase from the input mixture to synthesize two separated outputs. Open Unmix~\cite{stoter19} and MMDenseLSTM~\cite{takahashi2018mmdenselstm} are examples of spectrogram-based approaches, that show good results in the SiSEC 2018 evaluation~\cite{stoter20182018} on the MusDB dataset~\cite{MUSDB18}. 
Spleeter~\cite{hennequin2020spleeter} is another spectrogram-based system that has shown a strong performance and has been widely adopted by the digital music industry, but it is now outperformed by more advanced spectrogram domain architectures such as D3Net~\cite{takahashi2020d3net} and ByteMSS~\cite{kong2021decoupling}, even though Spleeter is trained on much more data from Bean dataset~\cite{pretet2019singing} with nearly 25,000 songs. 

The spectrogram-based separation methods often suffer from incorrect phase reconstruction. Kong et al.~\cite{kong2021decoupling}
proposed ByteMSS to estimate the phase by estimating complex ideal ratio masks where they decouple the estimation of these masks into magnitude and phase estimations based on
a deep residual U-Net model. This system achieves the state-of-the-art (SOTA) singing voice separation
result of SDR with 8.98 dB on the MUSDB18 test dataset. This opens doors to many other closely related MIR tasks, such as music transcription, singing melody transcription and lyrics transcription. 

In this paper, we explore end-to-end neural solutions to lyrics transcription that benefits from the best of both direct modeling and extraction-transcription studies. We explore the use of ByteMSS network for singing voice extraction.

\subsection{Lyrics Transcription of Vocal and Polyphonic Music}
The singing vocal carries the lyrics of a song, therefore lyrics transcription of polyphonic music can benefit a lot from the recent advances in the field of  lyrics transcription of solo-singing~\cite{gupta2022overview,gao2022self}.

\subsubsection{Lyrics Transcription of Solo-singing}
Lyrics transcription of solo-singing is typically performed by a speech recognition engine, such as Kaldi~\cite{povey2011kaldi}. Gupta et al. developed a DNN-HMM using an early version of DAMP dataset and reported 29.65\% word error rate (WER) with duration-adjusted phonetic lexicon. More recently, Dabike et al.~\cite{dabike2019automatic} created a cleaner version of DAMP dataset and constructed a factorized Time-Delay Neural Network (TDNN-F) using Kaldi~\cite{povey2011kaldi} with WER of 19.60\%. Demirel et al.~\cite{demirel2020automatic} further incorporated convolutional and self-attention layers to TDNN-F and achieved WER 14.96\%. Our recent work~\cite{gao2021tran} introduced an end-to-end Transformer-based framework along with RNN-based language model and is shown to be competitive with current approaches. This provides possibility and flexibility for lyrics transcription in polyphonic music where we would need a pre-trained solo-singing model to initialize the acoustic model for the integrated fine-tuning process. 

\begin{figure*}[t]
\centering
\includegraphics[width=178mm]{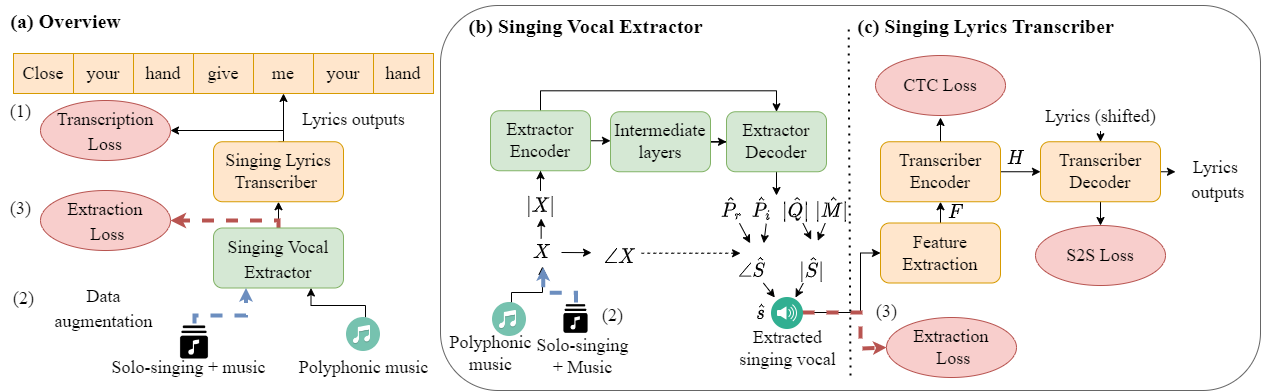}
\caption{(a) The overview network architecture of the proposed integrated fine-tuning approach (2) PoLyScriber with its two variants (1)(3). The detailed network architecture of (2) PoLyScriber includes (b) singing vocal extractor and (c) singing lyrics transcriber. PoLyScriber-NoAug (1) and PoLyScriber-L (3) have the same network architecture as PoLyScriber, while PoLyScriber-NoAug (1) does not employ data augmentation and PoLyScriber-L (3) has the additional extraction loss.}
\label{proposeFig}
\end{figure*}

\subsubsection{From Solo-Singing to Polyphonic Music} 
Studies show that acoustic models trained on solo-singing data \cite{mesaros2009adaptation,mesaros2013singing,kawai2016speech,kruspe2016bootstrapping, gupta2018automatic,tsai2018transcribing,gupta2018semi, dabike2019automatic} performs poorly on  polyphonic music test data \cite{dabikesheffield,guptalyricsAdapt,demirel2021low}. One way to adapt a solo-singing acoustic model towards polyphonic music is through domain adaptation~\cite{gupta2019acoustic}. It is found that an acoustic model adapted on polyphonic music data outperforms that adapted on extracted vocals. This suggests that singing voice extraction or separation introduces unwanted artifacts, that adversely affect the lyrics transcription performance. Despite much progress, the lyrics transcription of polyphonic music remains to be improved. As singing vocals and background music are highly correlated in polyphonic music, resulting in overlapping frequency components, lyrics transcription of polyphonic music is more challenging than that of solo-singing. 

\subsubsection{Towards End-to-End Singing Decoder}
Many lyrics transcription systems \cite{mesaros2010automatic,demirel2021low, demirel2020automatic,stoller2019,9054567,ahlback2021mstre,gao2022music} are based on hybrid modular architecture, that consists of acoustic, lexical and language models \cite{povey2011kaldi}, each with a different objective, thereby suffering from disjoint training issue. Studies show that end-to-end (E2E) systems bring us many advantages \cite{graves2014towards,miao2015eesen,chorowski2014end,chan2016listen,kim2017joint,nakatani2019improving} including its simpliciy without the need of elaborate controlling of GMM, HMM and neural network models, its independent capability of lexicon and its flexibility to incorporate other models. E2E model only needs one single neural network with one objective function to optimize for the lyrics transcription task. 
We reported an E2E transformer-based system for lyrics transcription of polyphonic music~\cite{gao2021tran} that outperforms other approaches in the literature. It consists of a transformer based encoder-decoder framework with a multi-head attention that implicitly learns the global time contextual dependency of the singing vocal utterance, beyond just the current frame. We adopt this singing decoder~\cite{gao2021tran} as our network backbone.

\section{Overview of PoLyScriber}
\label{PoLyScriber}
\subsection{Motivation}

\begin{table}[t]
\caption{Comparison of lyrics recognition (WER\%) performances under different conditions of polyphonic music testsets on the Solo-singing transcriber (SoloM) model as detailed in Section~\ref{Performance of the Pre-trained Models}. The extracted vocal and bgm are obtained by sRes Extractor as described in Section~\ref{Extractor} from polyphonic music testsets in Section~\ref{Polyphonic music dataset}. The lower the SDR is, the higher-level energy of the bgm are added.}
\footnotesize
\centering
\begin{tabular}{l|ccc}
\toprule
\textbf{Decoding input audio types to SoloM}    & \textbf{Hansen}   & \textbf{Jamendo}  & \textbf{Mauch}  \\ \midrule
extracted vocal              &    66.76     &   62.48 & 58.28    \\ 
extracted vocal + bgm (SDR = 20dB) & 59.33 &59.82 & 55.46\\

extracted vocal + bgm (SDR = 15 dB) & 58.67 & 58.72 & \textbf{54.67} \\
extracted vocal + bgm (SDR = 5 dB) & \textbf{56.09} &59.24&55.38\\
extracted vocal + bgm (SDR = 10 dB) &57.24& \textbf{58.49} &55.50\\
extracted vocal + bgm (SDR = 0 dB)      &     59.96  &   61.92 &61.62\\
extracted vocal + bgm (SDR = -5 dB)         & 67.77  &  69.16 & 74.20      \\ 
extracted vocal + bgm (SDR = -10 dB)    &  84.03   & 81.94 & 89.02 \\ 
extracted vocal + bgm (SDR = -15 dB)     &  93.37    & 90.70 & 95.66      \\ 
extracted vocal + bgm (SDR = -20 dB) & 97.10 &  94.80 & 96.70      \\ 
Original polyphonic tracks & 65.54 & 65.65 &69.43 \\
\bottomrule
\end{tabular}
\label{bgm}
\end{table}

Since singing vocals are highly correlated and overlapped with the background music in polyphonic music, it is difficult to achieve a perfect singing voice extraction. In many studies, singing extractors and singing acoustic modeling are two steps in a pipeline, where each of these modules is independently trained~\cite{mesaros2010automatic,demirel2021low, demirel2020automatic} and extracted vocals often suffer from distortions in the extractor. On the other hand, \cite{9054567,gao2022genre} have shown promising results where the vocal extraction step is avoided and the vocals along with the accompanying music are modeled together for lyrics recognition (i.e.~direct modeling). However, the music accompaniments can be loud which can affect lyrics intelligibility and hence the performance of the lyrics recognition system~\cite{9054567,gao2022genre,stoller2019}. 

We hypothesize that for lyrics recognition, the front end should be able to suppress loud background accompaniments while not causing vocal distortion or vocal removal. One way to test this theory is to add back the extracted background accompaniments (bgm) to the extracted vocal at different signal-to-distortion ratios (SDR) and pass this combined signal through a transcriber trained on singing vocals to observe its effect on the lyrics recognition performance. Indeed, we observe (Table~\ref{bgm}) that in the presence of low volume background accompaniments, lyrics recognition performance is better than the two extreme ends - extracted vocals input with completely suppressed background music, and the original vocals+music input. The reason is that extracted vocals with complete music suppression results in partial removal and distortions in the vocals while the original polyphonic music tracks have loud background music that affects lyrics recognition performance. Adding in some of the removed music also adds in the vocal parts, thus improving recognition performance.

This raises the question - How to automatically cause the ideal amount of music suppression required for lyrics recognition? 
To answer this question, in this work, we propose an integrated optimization strategy for the extractor and the acoustic model for the task of lyrics transcription.

\subsection{Integrated Fine-tuning Approach}
In this work, we propose an end-to-end lyrics transcription framework, PoLyScriber, in which we holistically train the singing voice extraction module and the singing acoustic model in a single integrated network utilizing polyphonic data augmentation for polyphonic music, as shown in~Fig.\ref{proposeFig}. To comprehensively explore the integrated fine-tuning approach, we also investigate two different variants of the PoLyScriber. These includes transcription-oriented training (PoLyScriber-NoAug) and the inclusion of the extraction loss (PoLyScriber-L), as illustrated in~Fig.\ref{proposeFig}.

\subsubsection{PoLyScriber with Polyphonic Data Augmentation}
PoLyScriber is an E2E approach utilizing polyphonic data augmentation where vocal extraction model is integrated with the lyrics transcription model and they are trained together solely towards the objective of lyrics transcription. This model uses both \textit{realistic} and \textit{simulated} polyphonic music data for E2E training, as shown in Fig.~\ref{proposeFig} (a) with dotted lines and the symbol (2).
Having access to solo-singing data, \textit{simulated} polyphonic music is created by data augmentation. 
We conduct polyphonic data augmentation by randomly selecting music and mixing it with solo-singing data with random signal-to-noise ratios (SNR) while training the PoLyScriber. By doing so, we expect to improve the diversity of polyphonic music data and further to be beneficial for model generalization. 

\subsubsection{PoLyScriber-NoAug Variant with Transcription-oriented Training}
If solo-singing or background music data is not available, we explore a transcription-oriented training strategy, PoLyScriber-NoAug, that does not adopt polyphonic data augmentation. PoLyScriber-NoAug has the same architecture as PoLyScriber trained solely towards the objective of lyrics transcription, while the training data used for PoLyScriber-NoAug is only \textit{realistic} polyphonic music data, as explained in Fig.\ref{proposeFig} (a) with symbol (1).

\subsubsection{PoLyScriber-L Variant with Inclusion of Extraction Loss}
On top of PoLyScriber, we also introduce a variant by additionally incorporating a vocal extraction loss along with the transcription loss to supervise the extractor front-end training on the \textit{simulated} as well as real data as illustrated in Fig. \ref{proposeFig}(a) with dotted lines and the symbol (3).

\subsection{Network Architecture}
We adopt the same network architecture for PoLyScriber and its variants. 
We simplify a residual-UNet vocal extraction front end~\cite{kong2021decoupling} as a simplified Residual-UNet (sRes) and concatenate it with a transformer-based singing acoustic model together to build PoLyScriber, and globally adjust the weights in each of these modules via back-propagation. 

To construct base models for integrated fine-tuning initialization, we first obtain the pre-trained singing vocal extractor and the singing acoustic model (transcriber). Specifically, the extractor front end is trained to reconstruct singing vocal using parallel polyphonic music and corresponding clean singing vocals data, and the transcriber is constructed with a joint encoder-decoder and connectionist temporal classification (CTC) architecture~\cite{nakatani2019improving} to transcribe lyrics from clean singing vocals.
Second, we use the pre-trained extractor and transcriber as the base models and further globally train the extractor and transcriber on a training dataset.
The integrated fine-tuning allows the linguistic information captured by the acoustic model to flow back and inform the extraction front end. By optimizing PoLyScriber network in this way, the extractor would be able to extract features in a way that is suitable for the task of lyrics transcription. 

\section{Components of PoLyScriber}
\label{PoLyScriber Components}
In this section, we describe our design of each of the components in the PoLyScriber in detail.

\subsection{Singing Vocal Extractor Front End}
\label{Singing Vocal Extractor Front-End}
The goal of the singing vocal extractor in PoLyScriber is to remove, or at least suppress the background music to a certain level while sustaining lyrics transcription related vocal parts. The extractor is based on the Residual-UNet~\cite{kong2021decoupling} model to estimate complex ideal ratio masks (cIRMs) and spectrogram, modified for the purpose of integrated fine-tuning. 
Compared with conventional approaches that do not estimate phases of the extracted signals~\cite{jansson2017singing,chandna2017monoaural,stoter2019open,takahashi2018mmdenselstm}, and often suffers from poor audio quality possibly with incorrect phase reconstruction, cIRMs based system contains phase prediction and therefore, may alleviate incorrect phase reconstruction problem. However, the direct predicting of real and imaginary parts of cIRMs suffers from being sensitive to time-domain signal shifts \cite{choi2018phase,wang2018supervised,tan2019complex,kong2021decoupling}.
Considering the superiority of cIRM for estimating phase \cite{kong2021decoupling}, we use the extractor that is designed to decouple the estimation of cIRMs into magnitude and phase estimations. The extractor also combines the predictions of the cIRM mask and spectrogram where the spectrogram prediction is the direct magnitude prediction serving as a residual component to complement the bounded mask prediction~\cite{kong2021decoupling}. This was one of the top performing systems in the recent Music Demixing Challenge 2021 \cite{mitsufuji2021music} with a signal-to-distortion ratio (SDR) of 8.9 dB on a standard test dataset.

We employ a simplified version of the Residual-UNet network as our front-end sRes vocal extractor that has an encoder, intermediate layers and a decoder as shown in Fig.~\ref{proposeFig} (b). We denote the cIRM as $\hat{M}$. The input short-time Fourier transforms (STFTs) $X$ from the realistic polyphonic music or the \textit{simulated} polyphonic music is compressed by the extractor encoder and intermediate layers into a lower dimensional descriptor and then the descriptor is re-expanded to the size of the target solo-singing STFTs $S$ by the extractor decoder. 
In the following subsections, we present the details of the vocal extractor.

\subsubsection{Extractor Architecture}
The extractor encoder consists of four residual encoder blocks (REBs) where each REB contains two residual convolutional blocks (RCBs) followed by an 2 $\times$ 2 average pooling layer to reduce the feature map size. 
The \textit{real}/\textit{simulated} polyphonic music STFT input $X$ is first encoded as polyphonic representations via the extractor encoder. Each RCB consists of two convolutional layers with kernel sizes 3 $\times$ 3. A residual connection is added between the input and the output of a RCB \cite{kong2021decoupling}. A batch normalization and a leaky ReLU non-linearity with a negative slope of 0.01 is applied before each convolutional layer inside each RCB following the pre-act residual network configuration \cite{kong2021decoupling}.

The two intermediate convolutional blocks (ICBs) then transform the polyphonic representations into a hidden descriptor, where each ICB has the same architecture as the REB except the pooling layer. The hidden descriptor is further fed into the extractor decoder, which is symmetric to those in the extractor encoder and it contains four residual decoder blocks (RDBs). 
Each RDB consists of a transposed convolutional layer with a kernel size 3 $\times$ 3 and stride 2 $\times$ 2 to upsample feature maps, followed by two RCBs. After the extractor decoder, an additional ICB and a final convolutional layer are applied to generate four outputs (the magnitude of the cIRM $|\hat{M}|$, direct magnitude prediction $|\hat{Q}|$, real part of the cIRM $\hat{P}_{r}$ and the imaginary part of the cIRM $\hat{P}_{i}$).
Therefore, the extracted singing vocal STFTs $\hat{S}$ is predicted following~\cite{kong2021decoupling} as below:

\begin{equation}
\begin{split} 
 \hat{S} =  |\hat{S}|cos\angle{\hat{S}} + j|\hat{S}|sin \angle\hat{S},\\
 |\hat{S}| = relu(|\hat{M}| |X| + |\hat{Q}|),\\
\angle{\hat{S}} = \angle{\hat{M}} + \angle{X},\\
cos\angle{\hat{S}} = cos\angle{\hat{M}}cos\angle{X}-sin \angle\hat{M}sin \angle X,\\
sin \angle\hat{S} = sin \angle X cos\angle\hat{M}+cos\angle X sin \angle \hat{M},\\
sin \angle \hat{M} = \hat{P}_{r}/\sqrt{\hat{P}_{r}^{2}+\hat{P}_{i}^{2}},\\
cos \angle \hat{M} = \hat{P}_{i}/\sqrt{\hat{P}_{r}^{2}+\hat{P}_{i}^{2}}
\end{split}
\label{extraction-eq}
\end{equation}
As illustrated in Fig.~\ref{proposeFig} (b) and Eq.~\ref{extraction-eq}, the angle of cIRM $\angle{\hat{M}}$ and the angle of extracted vocal STFTs $\angle{\hat{S}}$ can be obtained from $\hat{P}_{r}$ and $\hat{P}_{i}$, while $|\hat{Q}|$ is residual component to $|\hat{M}|$ for getting extracted vocal magnitude $|\hat{S}|$.
After obtaining the STFTs of the extracted vocal, an inverse STFT is applied to obtain the extracted vocal waveform.
\subsubsection{Extractor Training Objective}
The sRes extractor is pre-trained on parallel polyphonic music and clean solo-singing data with a L1-loss that is computed on the waveform domain as shown below:

\begin{equation}
\begin{split} 
 \mathcal{L}_{\text{Ext}} = \frac{1}{T}(\sum_{t=1}^{T}|\mathbf{\hat{s}}(t)-\mathbf{s}(t)|)
\end{split}
\end{equation}

where the $\mathbf{\hat{s}}(t)$ and $\mathbf{s}(t)$ are the extracted singing vocal waveform and the corresponding target solo-singing waveform with $t$ as the discrete time index, respectively.

\subsection{Singing Lyrics Transcriber}
\label{Singing Vocal Transcriber}
The singing lyrics transcriber uses a transformer-based lyrics transcription framework which is trained to decode input feature sequence of extracted singing vocal to output lyrical sequence, as shown in Fig.~\ref{proposeFig} (c).
The transcriber encoder converts the input acoustic features to intermediate representations, and the transcriber decoder predicts lyrical tokens (we use sub-words where 5,000 sub-words
are generated using byte-pair encoding (BPE) in this work) one at a time given the intermediate representations and previously predicted lyrical tokens in an auto-regressive manner.

\subsubsection{Transcriber Encoder}
\label{Poly-Transformer Encoder}
The transcriber encoder consists of an acoustic embedding module and twelve identical sub-encoders where each sub-encoder contains a multi-head attention (MHA)~\cite{vaswani2017attention} and a position-wise feed-forward network (FFN). 
The extracted vocal acoustic features $F$ for the transcriber are first obtained from the extracted singing vocal audio via the feature extraction block, which first performs downsampling on the audio to a sampling rate of 16KHz and then extracts 80-dim filterbank feature with a window of 25 ms, shifting 10 ms.
$F$ is then encoded into the acoustic embedding by the acoustic embedding module using subsampling and positional encoding (PE)~\cite{vaswani2017attention}. The acoustic embedding block contains two CNN blocks with a kernel size of 3 and a stride size of 2. The sub-encoders then transform the acoustic embedding into a hidden representation $H$.
Residual connection~\cite{he2016deep} and layer normalization~\cite{ba2016layer} are employed inside each of the sub-encoder. 

\begin{equation}
\centering
\begin{split}     
       H = \text{TranscriberEncoder}(F)
\end{split}
\end{equation}

\subsubsection{Transcriber Decoder}
The transcriber decoder consists of a lyrical embedding module and six identical sub-decoders, where each sub-decoder has a masked MHA, a MHA and a FFN. 
During training, $Y$ represents the lyrical token history that is offset right by one position, however, during run-time inference, it represents the previous predicted token history. $Y$ is first converted to lyrics token embedding via lyrical embedding module, that consists of an embedding layer, and a positional encoding (PE) operation.

\begin{equation}
\centering
\begin{split}    
       O = \text{TranscriberDecoder}(H,Y)
\end{split}
\end{equation}

\begin{table}[t]

\centering
\caption{A description of polyphonic music dataset that consists of DALI and NUS collections.}
\label{tab:datasets}
\footnotesize
\begin{tabular}{l|l|rrr}
\toprule
\multicolumn{2}{c}{\textbf{Name}}      & \textbf{\# songs} & \textbf{Lyrical lines} & \textbf{Duration} \\ \midrule
\multirow{2}{*}{Poly-train} & DALI-train & 3,913          & 180,034                & 208.6 hours       \\ 
                            & NUS        & 517            & 264,62                 & 27.0 hours        \\ \midrule
\multirow{2}{*}{Poly-dev}   & DALI-dev   & 100            & 5,356                  & 3.9 hours         \\ 
                            & NUS        & 70             & 2,220                  & 3.5 hours         \\ \midrule
\multirow{3}{*}{Poly-test}  & Hansen     & 10             & 212                    & 0.5 hour          \\ 
                            & Jamendo    & 20             & 374                    & 0.9 hour          \\ 
                            & Mauch      & 20             & 442                    & 1.0 hour          \\ \bottomrule
\end{tabular}
\end{table}

The lyrics embedding is fed into the masked MHA that ensures causality, i.e.~the predictions for current position only depends on the past positions. The output of the masked MHA and the acoustic encoding $H$ are then fed to the next MHA for capturing the relationship between acoustic information $H$ and textual information from the masked MHA. The residual connection~\cite{he2016deep} and layer normalization~\cite{ba2016layer} are also employed inside each of the sub-decoder. 
\subsubsection{Transcriber Learning Objective}
A combined CTC and sequence-to-sequence objective is conducted for the transcriber pre-training as shown below:
\begin{equation}
\begin{split} 
 \mathcal{L}_{\text{Transcriber}} = \alpha \mathcal{L}^{\text{CTC}} + (1-\alpha) \mathcal{L}^{\text{S2S}},\\
  \mathcal{L}^{\text{CTC}} = \text{Loss}_{\text{CTC}}(G_{ctc},R),\\
  \mathcal{L}^{\text{S2S}} = \text{Loss}_{\text{S2S}}(G_{s2s},R)   
\end{split}
\end{equation}
where $\alpha \in [0,1]$, $R$ is the ground-truth lyrical token sequence. The transcriber decoder are followed by a linear projection and softmax layers, that converts the decoder output $O$ into a posterior probability distribution of the predicted lyrical token sequence $G_{s2s}$. The sequence-to-sequence (S2S) loss is the cross-entropy of ground-truth lyrical token sequence $R$ and its corresponding predicted lyrical token sequence $G_{s2s}$. Also, a linear transform is applied on $H$ to obtain the token posterior distribution $G_{ctc}$. CTC loss is computed between $G_{ctc}$ and $R$~\cite{nakatani2019improving}.

\subsection{Learning Objective}
To leverage on the acoustic and linguistic knowledge from the pre-trained sRes extractor and transcriber models, PoLyScriber is initialized by the respective pre-trained models for training.
PoLyScriber is trained to minimize S2S and CTC losses with an objective function $\mathcal{L}_{\text{PoLyScriber}}$,
\begin{equation}
 \mathcal{L}_{\text{PoLyScriber}} = \alpha \mathcal{L}^{\text{CTC}} + (1-\alpha) \mathcal{L}^{\text{S2S}}
\end{equation}

During integrated fine-tuning, the parameters are differentiable and backpropogated all the way to the extractor towards the designed losses.
For the variants of PoLyScriber, PoLyScriber-NoAug shares the same objective function with PoLyScriber, while PoLyScriber-L is holistically optimized by S2S, CTC and extraction losses with an objective function $\mathcal{L}_{\text{PoLyScriber-L}}$,
\begin{equation}
 \mathcal{L}_{\text{PoLyScriber-L}} = \alpha \mathcal{L}^{\text{CTC}} + (1-\alpha) \mathcal{L}^{\text{S2S}} +  \mathcal{L}^{\text{Ext}}
 \end{equation}

At run-time inference, PoLyScriber and its variants directly converts the input polyphonic music to the output lyrical token sequence by passing through the globally trained network of the extractor and the transcriber. 

\section{Experiments}
\label{Experiments}

\subsection{Dataset}
\label{ssec:4.1}
Our experiments are conducted using four kinds of datasets - a polyphonic music dataset, a solo-singing dataset (clean vocals without background accompaniment), a music separation dataset, and a \textit{simulated} polyphonic music dataset.

\begin{table}[t]
\centering
\caption{A description of solo-singing dataset.}
\label{tab:datasets-mono}
\footnotesize
\begin{tabular}{l|rrr}
\toprule
\textbf{Name} & \textbf{\# songs} & \textbf{Lyrical lines}                                                   & \textbf{Duration} \\ \midrule
Solo-train         & 4,324       & 81,092 & 149.1 hours             \\ 
Solo-dev           & 66         &  482   & 0.7 hours               \\ 
Solo-test          & 70          &  480     & 0.8 hours               \\ \bottomrule
\end{tabular}

\end{table}
\subsubsection{Polyphonic Music Dataset}
\label{Polyphonic music dataset}
We explore the task of lyrics transcription on polyphonic music.
As shown in Table \ref{tab:datasets}, the polyphonic music training dataset, Poly-train, consists of the DALI-train~\cite{meseguer2018dali} dataset and a NUS proprietary collection. The DALI-train dataset consists of 3,913 English polyphonic audio tracks\footnote{There are a total of 5,358 audio tracks in DALI, but we only have access to  3,913 English audio links.}. The dataset is processed into 180,034 lyrics-transcribed audio lines with a total duration of 208.6 hours. The NUS collection dataset consists of 517 popular English songs. We obtain its line-level lyrics boundaries using the state-of-the-art audio-to-lyrics alignment system~\cite{9054567}, leading to 26,462 lyrics-transcribed audio lines with a total duration of 27.0 hours.

\begin{table*}[t]

\caption{A summary of the baselines and the proposed integrated fine-tuning models and their lyrics transcription results (WER\%) on both Poly-test and Poly-dev datasets.}

\footnotesize
\centering
\begin{threeparttable}
\begin{tabular}{lccccccccc}
\toprule
               &             & \multicolumn{1}{l}{}           & \textbf{} &  \textbf{Poly-dev} & \multicolumn{3}{c}{\textbf{Poly-test}}          \\ 
\textbf{Baseline Models}  &\textbf{Train data}  & \textbf{Extractor} & \textbf{Transcriber}   &     & \textbf{Hansen}         & \textbf{Jamendo}        & \textbf{Mauch}      \\ \midrule
pre-and-pre UMX~\cite{stoter2019open}  & Solo-train             & pre-train UMX           &   pre-train SoloM & 97.21   &97.87      &98.00 &97.59   \\
pre-and-pre sRes~\cite{kong2021decoupling}  & Solo-train             &pre-train sRes           &  pre-train SoloM &   67.77 &  66.76  &62.48 &58.28    \\
pre-and-fine UMX  & Poly-train         & pre-train UMX   &   fine-tune SoloM      &  63.06&   64.35    &  58.24     &  78.31 \\ 
pre-and-fine sRes & Poly-train         & pre-train sRes  &   fine-tune SoloM     &44.33 & 42.03      & 48.00        & 34.89      \\ 
Direct Modeling (DM)       & Poly-train        &  --    &      fine-tune SoloM            &   44.95    & 40.15       &       44.77&38.13  \\
DM-Aug        & Poly-train + DataAug        &  --    &   fine-tune SoloM           &       46.70   &35.61     &44.75       &36.43  \\
\midrule
\textbf{Integrated Fine-tuning Models}  &\textbf{Train data}  & \textbf{Extractor} & \textbf{Transcriber}   &  \textbf{Poly-dev}   & \textbf{Hansen}         & \textbf{Jamendo}        & \textbf{Mauch}      \\ \midrule
PoLyScriber-NoAug & Poly-train       &  fine-tune sRes   &   fine-tune SoloM     & 39.47  & 34.91 &    41.38   & 31.34         \\ 
PoLyScriber & Poly-train + DataAug      &   fine-tune sRes &   fine-tune SoloM      &  39.10 &\textbf{32.02}       &40.41   &\textbf{30.78}        \\ 
PoLyScriber-L & Poly-train + DataAug      &   fine-tune sRes  &  fine-tune SoloM       &  40.25  &   32.58       &\textbf{40.13}    &    32.77       \\ 
 \bottomrule

\end{tabular}
\begin{tablenotes}
\footnotesize
\item DataAug denotes the \textit{simulated} polyphonic music.
sRes and UMX denote simplified Residual-UNet and the UMX extractors, respectively.
\end{tablenotes}

\end{threeparttable}
\label{summary_table}
\end{table*}

The Poly-dev, serving as the validation dataset, consists of the DALI-dev dataset of 100 songs from DALI dataset \cite{9054567}, and 70 songs from a NUS proprietary collection. We adopt three widely used test sets -- Hansen\cite{hansen2012recognition},  Jamendo\cite{stoller2019}, and Mauch\cite{mauch2010lyrics} to form the Poly-test as shown in Table~\ref{tab:datasets}. The test datasets are English polyphonic songs, that are manually segmented into line-level audio snippets of an average length of 8.13 seconds, each of which we will refer to as an \textit{audio line}. In our experiments, we transcribe the lyrics of a song line-by-line following~\cite{gao2021tran}.

\subsubsection{Solo-singing Dataset}
We study the use of pre-training on solo-singing for lyrics transcription. A curated version~\cite{dabike2019automatic} of the English solo-singing dataset $\textit{Sing! 300} \times \textit{30} \times \textit{2}$~\footnote{The audio files can be accessed from https://ccrma.stanford.edu/damp/} is adopted, and detailed in Table~\ref{tab:datasets-mono}. A recent study~\cite{demirel2020automatic} reports the state-of-the-art performance on this dataset, that serves as a good performance reference. As indicated in~\cite{dabike2019automatic}, the training set Solo-train consists of 4,324 songs with 81,092 audio lines. The validation set Solo-dev and the test set Solo-test contain 66 songs and 70 songs with 482 and 480 audio lines respectively. The lyrics of this database are also manually transcribed~\cite{dabike2019automatic}.
\subsubsection{Music Separation Database}
We use a standard music separation database, MusDB18~\cite{MUSDB18}, for the singing vocal extractor pre-training and evaluation in PoLyScriber and its variants.
MusDB18 is a widely used database for singing voice separation and music source separation tasks, and it contains 150 full-length tracks with separate vocals and accompaniment where 86, 14 and 50 songs are designed for training, validation and testing, respectively. All songs are stereo with a sampling rate of 44.1 kHz, as described in~\cite{kong2021decoupling}. 

\subsubsection{\textit{Simulated} Polyphonic Music Dataset}
For the purpose of data augmentation for training the transcriber, PoLyScriber and PoLyScriber-L, we create a \textit{simulated} polyphonic music dataset. The \textit{simulated} training set (4,324 songs) is generated by adding music tracks, selected at random from MusDB18\cite{MUSDB18}, to every audio clip in Solo-train data at the time of training. Specifically, for each epoch of training, the solo-singing training set audio clips are mixed with random background accompaniment tracks at a wide range of signal-to-noise ratios (SNRs) sampled randomly between [-10dB, 20dB].

\subsection{Experimental Setup}

We detail the network architectures of the extractors and the transcribers as well as the training and decoding parameter settings for the proposed PoLyScriber and its variants as well as reference baselines in the following. Reference baselines in Table~\ref{summary_table} consist of the two-step pipelines (pre-and-pre and pre-and-fine strategies) and DM as described in Section~\ref{Singing Vocal Transcriber}.

\subsubsection{Extractor}
\label{Extractor}
To understand the effects of different extractor architectures on lyrics transcription, we implement two systems - the main extractor is the simplified Residual-UNet (sRes) described in Fig.~\ref{proposeFig} and Section~\ref{Singing Vocal Extractor Front-End}, which is one of the top performing systems for singing vocal extraction in Music Demixing Challenge 2021~\cite{mitsufuji2021music}, and another extractor for comparison is the Open-Unmix (UMX)~\cite{stoter19} that was the best performing open-source music source separation system in the source separation challenge SiSEC 2018~\cite{stoter20182018}.
We further detail the network architecture of the simplified Residual-UNet (sRes) and UMX below.

For Simplified Residual-UNet Vocal (sRes) Extractor, model compression is applied to reduce the model size and speed up the integrated fine-tuning process. The original Residual-UNet model \cite{kong2021decoupling} employs model architecture with a large parameter size of 102 M trainable parameters, which brings difficulties in model deployment due to the expensive and time-consuming computation. To reduce the model size, we apply model compression with several redundant layers removed while ensuring that the performance of vocal extraction does not get affected significantly.
Compared to the original Residual UNet \cite{kong2021decoupling}, in this work, we have removed two REBs, two ICBs and two RDBs where two RCBs are further removed in each REB and each RDB to establish the simplified Residual-UNet. The sRes extractor thus contains four REBs, two ICBs, and four RDBs with only 4.4 M trainable parameters.

Regarding to UMX Vocal Extractor, we utilize ``UMX" model from Open-unmix serving as UMX vocal extractor~\cite{stoter2019open} as a comparison system for the front end as it is a widely used open-source singing vocal extraction system.
UMX is trained by parallel polyphonic mixture and clean singing vocal audio in MusDB18 dataset~\cite{MUSDB18} using bidirectional Long short-term memory (BLSTM), and it learns to predict the magnitude spectrogram of singing vocal from the magnitude spectrogram of the corresponding mixture inputs (singing vocal+background music).
The network consists of a 3-layer BLSTM where each layer has 512 nodes. UMX is optimized in the magnitude spectrogram level using mean square error, and the singing vocal prediction is obtained by applying a mask on the input polyphonic music. 

We compare the two extractors briefly.
The sRes and UMX extractors are both spectrogram-based approach built on top of the MusDB18 dataset~\cite{MUSDB18} and produces the extracted vocal spectrogram, but the objective function of UMX is computed on spectrogram-level while that of sRes is on time-domain.
The idea of obtaining extracted vocal spectrogram are also different.
The sRes extractor estimates both phase and magnitude of the cIRM as well as incorporates the additional direct magnitude prediction to compensate the magnitude of cIRMs, while the UMX extractor does not predict the phases for extracted singing vocal that makes it suffer from incorrect phase reconstruction problem~\cite{kong2021decoupling}. 

\subsubsection{Baseline Frameworks}
Reference baselines are presented in Table~\ref{summary_table} and they include direct modeling (DM) as well as two-step pipeline approaches. Specifically for direct modeling method, DM and DM-Aug models are directly trained and validated on polyphonic music data, and they share the same architecture which only includes transformer-based transcriber that is pre-trained by solo-singing data.

There are two kinds of two-step pipeline strategies. First, in the pre-and-pre strategy \cite{dabike2019automatic,gupta2018semi}, the extractor and the transcriber are separately pre-trained. During inference, the trained extractor front end is used to get the extracted singing vocal that is then passed to the transcriber, that is trained on clean solo-singing vocals, for the decoding of the polyphonic music development and test sets. 
The second strategy, pre-and-fine, firstly employs pre-training on the front-end extractor for extracting vocals from polyphonic music, followed by fine-tuning solo-singing based transcriber using extracted vocal data \cite{gupta2019acoustic}. The solo-singing based transcriber is pre-trained by solo-singing data. Polyphonic music data also goes through the extractor for testing the pre-and-fine model. 

To be more specific, we use off-the-shelf SOTA systems for both the modules in our two-step baseline, where pre-and-pre and pre-and-fine approaches are experimented on two pre-trained extractors including simplified Residual-UNet and UMX. The pre-trained ``UMX''~\cite{stoter2019open} is utilized to extract singing vocals for the two-step pipelines pre-and-pre UMX and pre-and-fine UMX, and the pre-trained simplified Residual-UNet is employed to extract vocals for pre-and-pre sRes and pre-and-fine sRes models.
We note that the audio inputs for the feature extraction block are \textit{real}/\textit{simulated} polyphonic music for DM or DM-aug models, and extracted singing vocal for pre-and-pre and pre-and-fine models.

\subsubsection{Integrated Fine-tuning Frameworks}
Instead of directly training the transcriber on polyphonic music as in DM-aug, the proposed integrated fine-tuning approach alleviates the background music inference problem by integrating the extractor and the transcriber for training in a single network.
Specifically, we conduct pre-training of the simplified Residual-UNet extractor first, and then globally train the pre-trained extractor and transcriber in a single network. The pre-training uses MusDB18 dataset~\cite{MUSDB18} following the default parameter settings in~\cite{kong2021decoupling}, and the integrated fine-tuning of PoLyScriber and its variants (PoLyScriber-NoAug and PoLyScriber-L) uses the polyphonic music data as detailed in Table~\ref{summary_table}. Specifically, data augmentation is performed for PoLyScriber and PoLyScriber-L models but not for PoLyScriber-NoAug model. The feature extraction block inputs for PoLyScriber and its variants are intermediate extracted singing vocal audio from the extractors. 

\subsubsection{Training and Inference Configurations}
We use ESPnet \cite{watanabe2018espnet} with PyTorch back end to build all transcribers, and the interpolation factor $\alpha$ between CTC loss and S2S loss is set to 0.3 for training.
Other parameters of the transcriber follow the default setting in published LibriSpeech model (LS online)\footnote{See the pretrained librispeech model ``pytorch large Transformer with specaug (4 GPUs) + Large LSTM LM" from the ESPNET github\url{https://github.com/espnet/espnet/blob/master/egs/librispeech/asr1/RESULTS.md}.}, where attention dim is 512, the number of heads is 8 in MHA and FFN laryer dim is 2048.
During integrated fine-tuning, the extractor input are upsampled to 44.1k sampling rate using the training data as indicated in Table~\ref{summary_table}.
All models are trained using the Adam optimizer with a Noam learning rate decay as in \cite{vaswani2017attention}, 25,000 warmup steps, 2,000,000 batch-bin, and 20 epochs~\cite{vaswani2017attention}. 
We follow the default setting in ESPnet~\cite{watanabe2018espnet} to average the best 5 validated model checkpoints on the development set (Solo-dev for pre-and-pre models, librispeech dev dataset for LS model and Poly-dev for the rest models) to obtain the final model as in Table~\ref{summary_table}. 
We follow the common joint decoding approach~\cite{hori2018end,nakatani2019improving}, which takes CTC prediction score into account during decoding by setting different CTC weight. During decoding process of different models, the default setting is used where the penalty, beam width and CTC decoding weight are set to 0.0, 10 and 0.3, respectively.

Since singing vocals extracted from polyphonic music are noisy version of clean solo-singing vocals, it is reasonable to initialize the polyphonic lyrics transcriber with a pre-trained solo-singing lyrics transcription model. 
We developed the pre-trained solo-singing transcription model (SoloM) by first training a speech recognition model (LS) on LibriSpeech dataset \cite{panayotov2015librispeech} following LS online model setting with 80-dim fbank features. We note that the pre-trained LS and SoloM models have the same transformer-based E2E architecture design as in Section \ref{Singing Vocal Transcriber}.

\section{Results and Discussion}
\label{Results and Discussion}
We study the effects of pre-training, different extractor models, two-step strategies, integrated fine-tuning, data augmentation and the incorporation of the extraction loss on lyrics transcription of polyphonic music. In-depth spectrogram analysis, error analysis and music genre analysis are also performed to understand the behavior of lyrics transcription systems for different music genres. We further conduct ablation study on the proposed model to explore the contributions of each sub-component. Also, a comparative study of the proposed models with the existing approaches is presented. 

We present a summary of the performance of the baseline approaches - the two-step models, and direct modeling approaches, as well as the proposed approach and its variants in Table~\ref{summary_table}. We also present different integrated fine-tuning mechanisms that include transcription-oriented training and inclusion of the extraction loss.

\subsection{Evaluation Methods}
We report the lyrics transcription performance in terms of word error rate (WER), which is the ratio of the total number of insertions, substitutions, and deletions with respect to the total number of words. Signal-to-Distortion ratio (SDR) is used for evaluating the singing voice extraction following~\cite{stoter2019open,kong2021decoupling,mitsufuji2021music}, which is defined as: 
\begin{equation}
\label{SDR}
\centering
       SDR = 10log_{10}\frac{\sum_{n}||\mathbf{s}(n)||^{2} + \epsilon}{\sum_{n}||\mathbf{s}(n)-\hat{\mathbf{s}}(n)||^{2} +\epsilon},
\end{equation}
where $\textbf{s}(n) \in R$ denotes the waveform of the ground truth solo-singing and $\hat{\textbf{s}}(n)$ is the waveform of the estimate for the extracted singing vocal with n being the (discrete) time index. We use a small constant $\epsilon=10^{-7}$ in Eq.~\ref{SDR} to avoid divisions by zero. The higher the SDR score is, the better the output of the system is.
\subsection{Performance of the Pre-trained Models}
\label{Performance of the Pre-trained Models}

\begin{table}[t]
\caption{Comparison of speech recognition and lyrics recognition (WER\%) performances between published online models and our models.}
\footnotesize
\centering
\begin{tabular}{c|cc}
\toprule
\textbf{Pre-trained Speech Model}    & \textbf{LS-online}   & \textbf{LS}   \\ \midrule
dev-clean                 &    3.29      &    3.72    \\ 
dev-other                 &        8.49  &  9.91      \\ 
test-clean                &     3.63      &     4.05    \\ 
test-other                &     8.35    &        9.66 \\ \toprule
\textbf{Pre-trained Solo-sing Model} & \textbf{SOTA~\cite{demirel2020automatic}} & \textbf{SoloM} \\ \midrule
Solo-dev                  &  18.74   &       16.33        \\ 
Solo-test                 &     14.79  &          16.30   \\ \bottomrule
\end{tabular}
\label{pre-train}
\end{table}

We first present the pre-trained model performances of the extractor and the transcriber to prepare for constructing the proposed PoLyScriber.

As shown in Table~\ref{pre-train}, the LS model is comparable in performance with the online published LS-online model. The pre-trained solo-singing model SoloM is initialized with the LS model, and then trained on solo-singing database as detailed in Table~\ref{tab:datasets-mono}. In Table~\ref{pre-train}, we can see that the transformer-based SoloM model achieves competitive performance compared with the current published SOTA~\cite{demirel2020automatic} reference model that is based on Kaldi speech recognition engine. 

While developing the pre-trained model for the main extractor (we call it sRes in Table~\ref{summary_table}), we employ model compression on the original residual UNet~\cite{kong2021decoupling} with 102 M trainable parameters and obtain a simplified Residual-UNet with 4.4 M trainable parameters as described in Section~\ref{Extractor}. We observe that the simplified Residual-UNet (sRes) achieves comparable performance of SDR 8.41 dB on widely used MusDB18 testset \cite{MUSDB18} in comparison with the original model \cite{kong2021decoupling} of SDR 8.90 dB.

In Table~\ref{summary_table}, we also compare the two-step approaches on two different extractors, a popular open-source UMX and recent high performing vocal extractor Residual-UNet, eg. pre-and-fine sRes vs. pre-and-fine UMX. Note that the simplified Residual-UNet outperforms UMX for the task of singing vocal extraction on MusDB18 testset with SDR of 8.41dB and 6.32dB, respectively.
We observe a superior performance of the simplified Residual UNet extractor over the UMX extractor for both pre-and-pre and pre-and-fine approaches for lyrics transcription as well. This corroborates the intuition that an improved vocal extraction front end helps the downstream tasks, e.g. the lyrics transcription. To this regards, we perform the rest of the experiments using the simplified Residual-UNet extractor.

\subsection{Performance of the Baseline Approaches}
To explore the different strategies of incorporating singing vocal extractor for lyrics transcription, we report the lyrics transcription performances of two-step pipelines, i.e the pre-and-pre and pre-and-fine approaches. 

In Table~\ref{summary_table}, we can see that the pre-and-fine approaches significantly outperform the corresponding pre-and-pre approaches on both the extractors (eg. pre-and-fine sRes with WER of 34.89\% achieves relative 40.10\% improvement from pre-and-pre sRes with WER of 58.25\% on Mauch dataset). This consistent improvement from pre-and-pre approach to pre-and-fine approach indicates that pre-and-fine approach is more capable of addressing the mismatch problem that exists in the pre-and-pre approach between the training and testing for the transcriber.

We note that the DM approach outperforms pre-and-pre UMX, pre-and-pre sRes and pre-and-fine UMX systems across all datasets significantly, and performs better than pre-and-fine sRes for Hansen and Jamendo datasets. This suggests that the DM approach is as competitive as the two-steps pipelines, and the presence of music in DM actually contributes to compensate the distorted parts that is lack in two-step pipelines.

\subsection{Performance of the Integrated Fine-tuning Approaches}

\begin{figure*}[htbp]
\centering
\includegraphics[width=184mm]{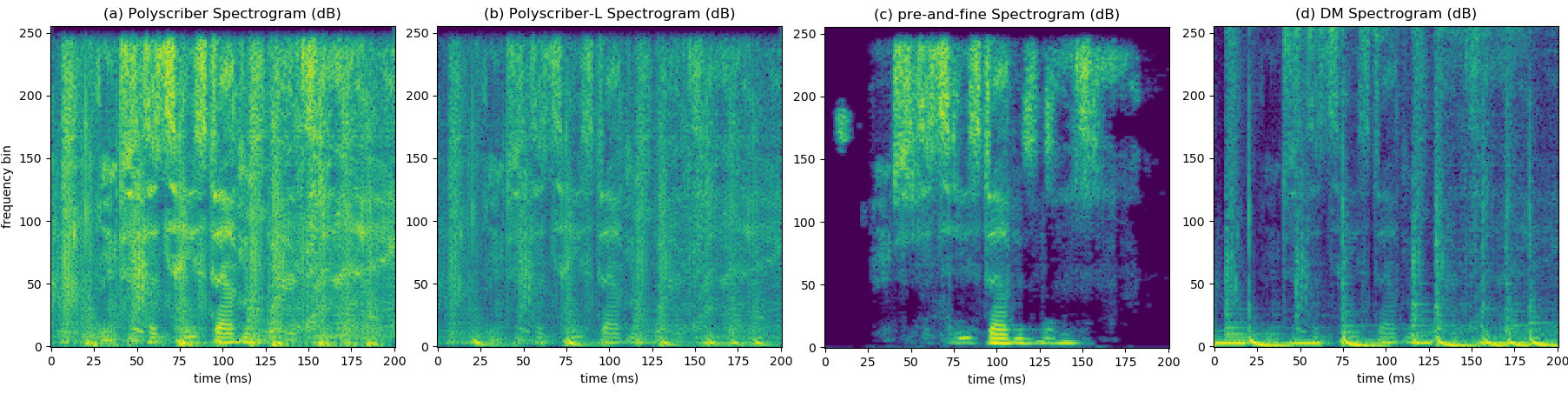}
\caption{The spectrograms of intermediate extracted vocals from (a) PoLyScriber extractor and (b) PoLyScriber-L extractor; (c) the spectrogram of pre-and-fine sRes model input vocal; and (d) the spectrogram of DM input original polyphonic music. The same 2-second audio clip is used for plotting spectrograms as the example in Table~\ref{example} with the transcription "Just a taste of my bad side".} 
\label{specFig}
\end{figure*}

\subsubsection{Comparison of Integrated Fine-tuning and Two-Step Pipelines}
We investigate the effect of the integrated fine-tuning by comparing the proposed PoLyScriber and its variants with the two-step approaches. In Table~\ref{summary_table}, we observe that PoLyScriber-NoAug framework consistently outperforms pre-and-fine sRes and pre-and-pre sRes models. This suggests that our integrated fine-tuning strategy of combining the training process of the extractor indeed yields better results in predicting lyrics compared to the two-step approaches, because both of these modules are optimized towards the common objective of lyrics transcription. Since the extractor in the two-step methods is not optimized towards the goal of lyrics transcription, the output from the extractor may not be a suitable input for the lyrics transcriber, leading to a sub-optimal solution. This is due to the fact that singing vocal extraction front ends are optimized to estimate the target singing vocal, which is different from optimizing for lyrics transcription. Therefore, the E2E PoLyScriber-NoAug, that optimizes the whole network towards lyrics transcription objective, performs better, and is able to address the mismatch problem between the front end and back end in the two-step approaches.

Furthermore, the superiority and effectiveness of the PoLyScriber-NoAug model over pre-and-fine sRes model suggests that the integrated fine-tuning approach is capable of performing effective transcription using only real-world polyphonic music data independently, without needing any parallel singing vocal and polyphonic music data generated from data augmentation. 

\begin{table*}[t]
\centering
\caption{Examples of transcribed lyrics with five representative models. \# C I D S denote the number of correct words (C), substitutions (S), insertions (I) and deletions (D).}
\footnotesize
\begin{tabular}{l|l|rc}
\toprule
 & \begin{tabular}[c]{@{}l@{}}\textbf{reference (ref)}: get a taste of my bad side just a taste of my bad side just a taste of my bad side\end{tabular}   &\# C I D S & WER (\%) \\ \midrule
\textbf{pre-and-fine sRes}       & \begin{tabular}[c]{@{}l@{}}\textbf{hypothesis (hyp)}: get your text up and touch the taste of \end{tabular}                                              & 3  0 12 6 &   85.71 \\ \hline
\textbf{DM}                & \begin{tabular}[c]{@{}l@{}}\textbf{hyp}: get a taste of my bad side Im just a taste of my bad side\end{tabular}                                        &   11 0 6 4 &  47.62  \\ \hline
\textbf{PoLyScriber-NoAug} & \begin{tabular}[c]{@{}l@{}}\textbf{hyp}: get a taste of my bad time just a taste of my bad side\end{tabular} & 12 0 7 2  &  42.86   \\ \hline
\textbf{PoLyScriber}       & \begin{tabular}[c]{@{}l@{}}\textbf{hyp}: get a taste of my bad times into the taste of my body just a taste of my\end{tabular}       &14 0 3 4  &   33.33  \\ \hline
\textbf{PoLyScriber-L}       & \begin{tabular}[c]{@{}l@{}}\textbf{hyp}: get a taste of my outside and just a taste of my bad time just a taste of my\end{tabular}       &      16 0 2 3 &   23.81 \\\bottomrule
\end{tabular}
\label{example}
\end{table*}

\subsubsection{Comparison of Integrated Fine-tuning and Direct Modeling}
To study the importance of tackling music interference problem for lyrics transcription, we compare direct modeling (DM) approach with integrated fine-tuning approach. DM method directly trains the transcriber on polyphonic music data without considering the background music as an interference. In Table~\ref{summary_table}, we observed that PoLyScriber-NoAug outperforms the DM method across all the datasets, which suggests that lyrics transcription can benefit from the integrated fine-tuning process with the extractor to handle the music interference problem. 

It is interesting to note that the DM method has a comparable, sometimes even better, performance than the two-step methods. This has also been observed previously in \cite{guptalyricsAdapt,9054567}. This may be an indication that background music may not be as harmful as the vocal artifacts introduced by the vocal extractor in lyrics transcription  \cite{9054567,gao2021tran}. This observation also indicates that the integrated fine-tuning would finally optimize the extractor into a state where its output will be somewhere in-between absolute background music suppression and complete background music presence, so as to gain from the benefits of both.

\begin{table*}[t]
\centering
\caption{Analysis of lyrics recognition performance in terms of errors (substitutions (S), insertions (I) and deletions (D).) for Poly-test. These are the number of words in error and the percentage of each error with respect to the total number of words.}
\footnotesize
\begin{tabular}{c|ccc|ccc|ccc|ccc|ccc}
\toprule
 & \multicolumn{3}{c|}{\textbf{pre-and-fine sRes}} & \multicolumn{3}{c|}{\textbf{DM}} & \multicolumn{3}{c|}{\textbf{PoLyScriber-NoAug}} & \multicolumn{3}{c|}{\textbf{PoLyScriber}}& \multicolumn{3}{c}{\textbf{PoLyScriber-L}} \\\midrule
Poly-test & \# I  & \# D &\# S &\# I& \# D & \# D& \# I &\# D &\# S&  \# I& \# D & \# S&  \# I& \# D & \# S\\ \midrule
All & 22&248 &229 &24& 188  &253&24&182 &225&35& 115 &257   &36 &\textbf{104} &247 \\
All /\ \# total words (\%) &  0.16 & 1.87&  1.73  & 0.18 & 1.42 &  1.91 &0.19 &1.37  & 1.70  &0.26 & 0.87&  1.94    &0.28& \textbf{0.78} &  1.87  \\
\bottomrule
\end{tabular}
\label{all_error_analysis}
\end{table*}

\subsubsection{Data Augmentation}
Data augmentation is employed by randomly adding music to solo-singing to create \textit{simulated} polyphonic music data while training the transcriber. We use this data augmentation method in direct modeling and integrated fine-tuning approaches. In Table~\ref{summary_table}, we can see that DM-Aug outperforms DM for all the three testsets, and PoLyScriber outperforms PoLyScriber-NoAug. This indicates that the proposed singing data augmentation is a simple yet beneficial solution to the problem of lack of diversity in training data, thereby improving model generalization.

\subsubsection{The Inclusion of the Extraction Loss}
Having access to the parallel solo-singing and \textit{simulated} polyphonic music, we are able to incorporate the extraction loss for the integrated fine-tuning framework. We investigate if the additional extraction loss is helpful to the integrated fine-tuning. In Table~\ref{summary_table}, we can see that the incorporation of the extraction loss in the PoLyScriber-L model shows a slight performance degradation on hansen and mauch datasets. Similar observations have been made in ASR studies where ASR-oriented optimization is effective for multi-talker speech recognition but an additional separation loss would not bring improvement to the recognition performance~\cite{wang2016joint,chang2019mimo,wu2021investigation}.

To analyze the relationship between the transcriber and the extractor in the integrated fine-tuning approaches, we present the extracted singing vocals from different systems and the corresponding original polyphonic music, along with their predicted transcriptions at the link~\footnote{\url{https://xiaoxue1117.github.io/PaperSample/}}. 
Qualitatively, when we listened to the outputs of the extractor after integrated fine-tuning, we found that some background music/noise got added back into the audio and distorted vocal parts in two-step approaches were recovered in integrated fine-tuning approach. Some interesting observations were that the lyrics transcription improved, compared to two-step methods, especially in the chorus sections of the song where multiple voices were present simultaneously. The integrated fine-tuning seemed to have drowned out the other vocals present in the background by adding back some music/noises. 
This gives evidence that the proposed integrated network adds in some background
music/noise and settles somewhere between completely suppressed background music (as in two-step approaches) and completely present background music (as in DM approaches), yielding better results than both.
Similar observation are also presented through spectrogram visualization in the following Section \ref{error and spectrogram}. 

\begin{table}[t]
\vspace{-0.5cm}
\centering
\caption{The genre distribution for polyphonic music Dataset, genre-specific average word duration per utterance and lyrics transcription performance (WER\%) on polyphonic music test sets.}
\footnotesize
\begin{tabular}{l|ccc}
\toprule
    \textbf{Statistics}  & \textbf{Metal}  & \textbf{Pop}   & \textbf{Hip hop} \\ \midrule
          \# songs Hansen &  3&6  & 1 \\ 
     \# songs Jamendo & 7 &9  &4  \\ 
          \# songs Mauch &8  &12  &0  \\ \midrule
      Percentage in Poly-train & 35\% & 59\% & 6\% \\ 
            Percentage in Poly-dev & 48\% & 49\% & 3\% \\ 
            Percentage in Poly-test & 34\% & 56\% & 10\% \\ 
\midrule
  Word duration in Poly-train (seconds)& 0.96& 0.90 & 0.65 \\ 
Word duration in Poly-dev (seconds) &0.76 &0.69  &0.58  \\ 
Word duration in Poly-test (seconds) &0.78 & 0.74 & 0.44 \\ \midrule
    \textbf{Model}  & \textbf{Metal}  & \textbf{Pop}   & \textbf{Hip hop} \\ \midrule
pre-and-fine UMX & 75.55 &64.43  &69.44 \\
pre-and-fine sRes & 49.83 & 35.31 &60.85 \\
             DM & 52.17 &36.85  &52.69 \\
 DM-Aug &  51.73 &35.55  &50.39 \\
PoLyScriber-NoAug-Pop  & 46.47 &34.53 &54.04\\
PoLyScriber-NoAug-Hiphop   &63.85  &47.18 &59.16\\
PoLyScriber-NoAug-Metal & 47.56 &35.68 &58.46\\
  PoLyScriber-NoAug  &\textbf{43.70}  &32.40  &51.54\\
    PoLyScriber  & 46.16  &\textbf{31.20}  &\textbf{46.55}\\
        PoLyScriber-L  & 46.79 &32.07  &46.88\\
\bottomrule
\end{tabular}
\label{genre}
\end{table}

\subsection{Error Analysis and Spectrogram Visualization}
\label{error and spectrogram}

We show an example in Table~\ref{example} and conduct an error analysis on the Poly-test in Table~\ref{all_error_analysis}.
We also visualize the magnitude spectrogram of a 2-second snippet for the example in Fig.~\ref{specFig}.
It is noted that the number of deletions in the proposed PoLyScriber and its variants is substantially reduced from that of the pre-and-fine and DM approaches. Furthermore,  PoLyScriber outperforms PoLyScriber-NoAug, which suggests the integrated fine-tuning helps to recover words that are deleted by the traditional methods. This again confirms the effectiveness of the integrated fine-tuning and data augmentation. 

We can observe from Fig.~\ref{specFig} that the DM input spectrogram Fig.~\ref{specFig} (d) contains loud music accompaniments and pre-and-fine sRes input spectrogram Fig. 2 (c) suffers from the vocal distortion especially for the duration of 0s-0.75s and 1.5s-2s, which has been alleviated by the proposed integrated fine-tuning approach as in Fig.~\ref{specFig} (a) (b).
By comparing Fig.~\ref{specFig} (d) with Fig.~\ref{specFig} (a) and Fig.~\ref{specFig} (b), we observe that the background accompaniments are removed from the spectrograms by the PoLyScriber and its variants, leading to improved lyrics transcription. The audio samples are available online~\footnote{\url{https://xiaoxue1117.github.io/PaperSample/}}.

\subsection{Music Genre Analysis}

We analyse the performance of different music genres per song on the polyphonic test sets -- Hansen, Jamendo and Mauch. The music genre distribution and average lyrical word duration per utterance in poly-train, poly-dev and poly-test is summarized in Table~\ref{genre}, and the test datasets consist of three broad genre categories -- pop, hip hop and metal, as given in \cite{9054567}. We report the lyrics transcription performance by music genre for our best performing models in Table~\ref{genre}.

We observe that the proposed PoLyScriber and its variants consistently outperform the DM-based approaches and two-step approaches across all the genres, which further verifies the effectiveness of the integrated fine-tuning technique over the conventional approaches. PoLyScriber-NoAug model without data augmentation performs the best for the metal songs, which suggests that the SNR-based diversity generated in the training data through our augmentation method does not appropriately generalize the model for metal songs. Metal songs often contain loud distorted guitar sounds which is not that common in pop and hip-hop songs. The kinds of instruments used in metal songs differ from pop or hip-hop songs, and the music tracks used for augmentation were mostly from pop genre, which may have been the reason of this drop in performance for metal songs. 
For hip-hop songs, data augmentation involved models including PoLyScriber and PoLyScriber-L perform better than other models without data augmentation. This suggests that data augmentation provides more diversity and robustness that helps in lyrics transcription of hip-hop songs.

\begin{table}[t]
\vspace{-0.5cm}
\centering
\caption{The genre distribution for MusDB18 test set and vocal extractor performance (SDR dB) on MusDB18 test set.}
\footnotesize
\begin{tabular}{l|ccc}
\toprule
    \textbf{Model}  & \textbf{Metal}  & \textbf{Pop}   & \textbf{Hip hop}  \\ \midrule
     PoLyScriber-NoAug  & 1.383 &1.565   &1.419\\
 PoLyScriber-NoAug-Pop  &  1.502&1.613&1.073\\
  PoLyScriber-NoAug-Hiphop   &5.320 &4.231 &6.438\\
    PoLyScriber-NoAug-Metal & 2.783&3.145 &3.448\\
    PoLyScriber  & -0.983 &-0.876 &-1.783 \\
    PoLyScriber-L  &0.547 & 0.557&0.106\\
        \toprule
    \textbf{Statistics} & \textbf{Metal}  & \textbf{Pop}   & \textbf{Hip hop}\\ \midrule
         \# songs in MusDB18 testset & 33 &15 & 2 \\ 
\bottomrule
\end{tabular}
\label{genreSDR}
\end{table}

\begin{table*}[t]
 \caption{Ablation study of PoLyScriber-L on lyrics transcription performance (WER\%) of polyphonic music.}
\centering
\footnotesize
\begin{tabular}{l|cc|cccc}
\toprule
\textbf{Ablation Model} &\textbf{Extractor}&\textbf{Transcriber} &\textbf{Poly-dev} & \textbf{Hansen} & \textbf{Jamendo} & \textbf{Mauch} \\ \midrule
PoLyScriber-L      & fine-tune & fine-tune   &  40.25  &   32.58       &\textbf{40.13}    &    32.77   \\ \midrule
-ExtLoss (PoLyScriber) & fine-tune & fine-tune & \textbf{39.10}  &  \textbf{32.02}  &      40.41  &   \textbf{30.78}            \\ 
-ExtLoss, DataAug (PoLyScriber-NoAug) & fine-tune& fine-tune& 39.47  & 34.91 &    41.38   & 31.34\\
-ExtLoss, DataAug, Pre-train extractor (model A) 
& scratch & fine-tune& 42.75 & 39.38 & 44.06    &44.27 \\
-ExtLoss, DataAug, Pre-train extractor-transcriber (model B)
& scratch & scratch&  55.80& 62.82 &65.49   &67.37 \\
-ExtLoss, DataAug, fine-tune extractor (pre-and-fine sRes)& fix& fine-tune &44.33 & 42.03      & 48.00        & 34.89             \\ 
-ExtLoss, DataAug, Extractor (DM) & remove& fine-tune &   44.95    & 40.15       &       44.77&38.13     \\
-ExtLoss, DataAug, Extractor, fine-tune transcriber (DM-Scratch) & remove& scratch &   53.52   &   57.66     &   65.68  &    48.17   \\
\bottomrule
\end{tabular}
\label{ablation}
\begin{tablenotes}
\footnotesize
\item[1] Notes: - denotes the removal of certain components from the PoLyScriber-L model.
\end{tablenotes}
\end{table*}

Furthermore, we observe from Table~\ref{genre} that metal songs have the longest lyrical word duration, and lyrical word duration for hip-hop songs is significantly lower than that of metal and pop songs across all datasets. This indicates that the hip-hop songs show higher syllable rates, and genre affects lyrical words in terms of speaking rate.
As described in~\cite{condit2015catching,9054567}, metal and hip-hop songs show lower lyrics intelligence compared with pop songs. Specifically, metal songs possess louder background music than pop songs and "Death Metal" songs shows zero lyrics intelligibility score~\cite{condit2015catching}. Hip-hop songs that always consist of rap with many words have a higher syllable rate and rapid vocalization than pop songs, thereby receiving lower lyrics intelligibility scores~\cite{condit2015catching,9054567}. Moreover, the percentage of available training data for pop songs is much more than metal and hip-hop songs. This explains why hip-hop and metal songs have a higher word error rate than pop songs across all models in Table~\ref{genre}.

To better understand model generalization ability across genres for both extractor and transcriber, genre-specific integrated fine-tuning (NoAug) models are trained using genre-specific data from Poly-train while tested on all the genres. For example, the PoLyScriber-NoAug-Pop model is trained on pop songs from Poly-train and tested on all the genres from Poly-test. We present extractor performance on MusDB18 test set together with its music genre distribution in Table~\ref{genreSDR} and transcriber performance in Table~\ref{genre}. We observe that the model with the extraction loss (PoLyScriber-L) generally performs better than the model without the extraction loss (PoLyScriber) on the singing vocal extraction task. This indicates that the inclusion of the extraction loss to integrated fine-tuning is beneficial to singing vocal extraction task. Table VI shows that PoLyScriber-NoAug-Hiphop for extraction purpose generalizes well on metal and pop songs but it performs not well on transcription performance in Table~\ref{genre}, and other genre-specific models also show good generalization ability on other genres that are not covered in the training set. This indicates that there is a tradeoff between extractor and transcriber for PoLyScriber-NoAug-Hiphop, and it is not well-trained towards lyrics transcription purpose with less training data while it remains the pre-trained vocal extractor performance. We can observe from Table~\ref{genre} that genre-specific models show worse generalization ability than the proposed model and its variants in terms of lyrics transcription performance, while genre-specific models perform better than the proposed models for vocal extraction in Table~\ref{genreSDR}. This indicates that there is a trade-off between the extractor and the transcriber performances while models converge, wherein transcription performance improves when the extractor performs worse.

\begin{table}[t]
\vspace{-0.8cm}
\caption{Comparison between the proposed solutions and other existing competitive solutions on whole song test to lyrics transcription (WER\%) of polyphonic music.}
\footnotesize
\centering
\begin{tabular}{c|cccc}
\toprule
\textbf{Kaldi-based Models}        & \textbf{Hansen} & \textbf{Jamendo} & \textbf{Mauch}& \textbf{Dali-test} \\ \midrule

RB1~\cite{dabikesheffield}                         & 83.43           & 86.70            & 84.98      & -       \\ 
DDA2~\cite{demirel2020recursive}                         & 74.81           & 72.15            & 75.39     & -        \\ 
DDA3~\cite{demirel2020recursive}                         & 77.36           & 73.09            & 80.66       & -      \\ 
CG~\cite{9054567}                          & -               & 59.60            & 44.00     & -        \\ 
GGL2~\cite{gaolyrics}                         & 48.11           & 61.22            & 45.35   & -          \\ 
GGL1~\cite{gaolyrics}                         & 45.87           & 56.76            & 43.76     & -        \\
MSTRE-Net \cite{ahlback2021mstre}        &  36.78       &   \textbf{34.94}  &    37.33&    42.11   \\
\midrule
\textbf{End-to-End Models}         & \textbf{Hansen} & \textbf{Jamendo} & \textbf{Mauch}& \textbf{Dali-test} \\ \midrule
DS~\cite{stoller2019}                          & -               & 77.80            & 70.90      & -       \\ 
E2E Transformer \cite{gao2021tran}    &   40.02  &    45.19    &  38.96&   46.56   \\
Multi-transcriber \cite{gao2021tran}    &   36.85    &   44.12     &33.69   & 40.20      \\
PoLyScriber-NoAug (ours)  & 35.77 &    41.10&  33.39 &   40.45     \\ 
PoLyScriber (ours)  &  33.57  &    41.32  &\textbf{32.83}  &         \textbf{36.52} \\ 
PoLyScriber-L (ours) &\textbf{33.46}    &41.00   &34.87   &   37.01 \\ 
\bottomrule
\end{tabular}
\label{existing}
\end{table}

\subsection{Ablation Study}
To study where the contributions come from and verify the effectiveness of the integrated fine-tuning, we conduct an ablation study on the integrated fine-tuning model with extraction loss (PoLyScriber-L) in Table~\ref{ablation}.
The removal of extraction loss brings slight improvement for lyrics transcription performance, which further verifies the advantage of transcription-oriented optimization for integrated fine-tuning. 
Moreover, we wonder if we can train a singing vocal extractor without parallel vocal and mixture data via integrated fine-tuning system. And can we even train the system from scratch?

A clear benefit of fine-tuning the extractor can be noticed while removing the extractor fine-tuning of PoLyScriber-NoAug. Concerning the lack of the extractor pre-training, model A still outperforms pre-and-fine sRes and DM for two testsets. To this regards, the performance of model A is acceptable and it is possible to train a singing vocal extractor without parallel vocal and mixture data inside the integrated fine-tuning framework. We can further observe that the model B suffers from a big performance drop from model A, which shows it is essential to pre-train the transcriber.

\begin{table}[t]
\vspace{-0.8cm}
\caption{Comparison of lyrics transcription (WER\%) of polyphonic music on line-level testsets}
\footnotesize
\centering
\begin{tabular}{l|lll}
\toprule
\textbf{Line-level test} & \textbf{Hansen} & \textbf{Jamendo} & \textbf{Mauch}\\ \midrule
E2E Transformer~\cite{gao2021tran}                    & 39.87 & 44.26  & 36.80\\ 
Multi-transcriber \cite{gao2021tran}     & 36.34  & 43.44   & 31.82\\ 
PoLyScriber-NoAug (ours)  & 34.91 &    41.38 & 31.34      \\ 
PoLyScriber (ours)  &  \textbf{32.02}  &      40.41     &   \textbf{30.78}      \\ 
PoLyScriber-L (ours)  &32.58       &\textbf{40.13}    &    32.77    \\ 
\bottomrule
\end{tabular}
\label{segmemt}
\end{table}

To compare the two-step pipelines with integrated fine-tuning, we can find that the lyrics transcription performance drops significantly from integrated fine-tuning approach (PoLyScriber-NoAug) to pre-and-fine sRes and DM. This indicates that integrated fine-tuning is critical to improve lyrics transcription performance. We also notice that the pre-training for the transcriber is also important while comparing DM with DM-Scratch.
\subsection{Comparison with the State-of-the-Art}
We compare PoLyScriber and its variants with the existing approaches for lyrics transcription on the polyphonic music testsets in Table~\ref{existing}. Specifically, we would like to compare PoLyScriber and its variants with the SOTA reference models~\cite{stoller2019,9054567,dabikesheffield,demirel2020recursive,gaolyrics,ahlback2021mstre}. Stoller et al.'s \cite{stoller2019} system is based on E2E Wave-U-Net framework and ~\cite{gao2021tran} is built on an E2E transformer with multi-transcriber solutions. The rest of the systems~\cite{9054567,dabikesheffield,demirel2020recursive,gaolyrics,ahlback2021mstre} are all based on the traditional Kaldi-based ASR approach. 

In Table \ref{existing}, we first report the lyrics transcription performance of the existing systems on the same test sets for whole songs evaluation. We decode short nonoverlapping segments of songs in Hansen, Jamendo and Mauch using our proposed models and combine the transcriptions of these segments to report WER results for the complete songs as in \cite{gao2021tran}. We also test on a larger database DALI-test \cite{ahlback2021mstre} with 240 whole-song polyphonic recordings processed as in \cite{gao2021tran}. We observe that the proposed PoLyScriber outperforms all previous E2E and Kaldi-based approaches for Hansen, Mauch and the large set DALI-test, which shows the general superiority of the integrated fine-tuning model over the conventional pipelines.
To avoid segmentation problems, we further consider the performance of E2E models on line-level test sets for comparison as in \cite{gao2021tran} in Table~\ref{segmemt}, which shows that the proposed PoLyScriber and its variants outperform the E2E models~\cite{gao2021tran}, thereby achieving better performance among all the existing models. We also note that the PoLyScriber-NoAug without data augmentation performs better than E2E models, which shows the proposed integrated fine-tuning is capable of achieving good lyrics transcription performance without the need of data augmentation where the parallel solo-singing and polyphonic music is created.
\section{Conclusion}

\label{Conclusion}
PoLyScriber serves as an important step into the exploration of lyrics transcription for polyphonic music via an E2E integrated fine-tuning setting.
We propose and idea to globally train singing voice extraction with lyrics transcriber towards the lyrics transcription objective for polyphonic music for the first time. 
We advocate the novel PoLyScriber framework with data augmentation for lyrics transcription of polyphonic music, that is proven to be effective. The proposed data augmentation paradigm enables PoLyScriber to leverage diverse polyphonic pattern and music knowledge from \textit{simulated} polyphonic music for hip-hop songs. We have shown that the proposed PoLyScriber and its variants outperform baseline frameworks for lyrics transcription through a comprehensive set of experiments on publicly available datasets.

\ifCLASSOPTIONcaptionsoff
  \newpage
\fi

\bibliographystyle{IEEEbib}
\bibliography{strings}

\begin{thebibliography}{10}

\bibitem{good2015efficacy}
A.~J. Good, F.~A. Russo, and J.~Sullivan,
\newblock ``The efficacy of singing in foreign-language learning,''
\newblock {\em Psychology of Music}, vol. 43, no. 5, pp. 627--640, 2015.

\bibitem{zhao2018sound}
Hang Zhao, Chuang Gan, Andrew Rouditchenko, Carl Vondrick, Josh McDermott, and
  Antonio Torralba,
\newblock ``The sound of pixels,''
\newblock in {\em Proceedings of the European conference on computer vision},
  2018, pp. 570--586.

\bibitem{zhao2019sound}
Hang Zhao, Chuang Gan, Wei-Chiu Ma, and Antonio Torralba,
\newblock ``The sound of motions,''
\newblock in {\em Proceedings of the IEEE/CVF International Conference on
  Computer Vision}, 2019, pp. 1735--1744.

\bibitem{gan2020music}
Chuang Gan, Deng Huang, Hang Zhao, Joshua~B Tenenbaum, and Antonio Torralba,
\newblock ``Music gesture for visual sound separation,''
\newblock in {\em Proceedings of the IEEE/CVF Conference on Computer Vision and
  Pattern Recognition}, 2020, pp. 10478--10487.

\bibitem{gan2020foley}
Chuang Gan, Deng Huang, Peihao Chen, Joshua~B Tenenbaum, and Antonio Torralba,
\newblock ``Foley music: Learning to generate music from videos,''
\newblock in {\em European Conference on Computer Vision}. Springer, 2020, pp.
  758--775.

\bibitem{chen2022sound2synth}
Zui Chen, Yansen Jing, Shengcheng Yuan, Yifei Xu, Jian Wu, and Hang Zhao,
\newblock ``Sound2synth: Interpreting sound via fm synthesizer parameters
  estimation,''
\newblock {\em Proc of the Thirty-First International Joint Conference on
  Artificial Intelligence AI and Arts}, pp. 4921--4928, 2022.

\bibitem{collister2008comparison}
LAUREN~B COLLISTER and DAVID HURON,
\newblock ``Comparison of word intelligibility in spoken and sung phrases,''
\newblock {\em Empirical Musicology Review}, vol. 3, no. 3, 2008.

\bibitem{dzhambazov2017knowledge}
G.~Dzhambazov,
\newblock {\em Knowledge-based Probabilistic Modeling for Tracking Lyrics in
  Music Audio Signals},
\newblock Ph.D. thesis, Universitat Pompeu Fabra, 2017.

\bibitem{hosoya2005lyrics}
T.~Hosoya, M.~Suzuki, A.~Ito, S.~Makino, L.~A. Smith, D.~Bainbridge, and I.~H.
  Witten,
\newblock ``Lyrics recognition from a singing voice based on finite state
  automaton for music information retrieval.,''
\newblock in {\em ISMIR}, 2005, pp. 532--535.

\bibitem{fine2014making}
Philip~A Fine and Jane Ginsborg,
\newblock ``Making myself understood: perceived factors affecting the
  intelligibility of sung text,''
\newblock {\em Frontiers in psychology}, vol. 5, pp. 809, 2014.

\bibitem{fujihara2012lyrics}
Hiromasa Fujihara and Masataka Goto,
\newblock ``Lyrics-to-audio alignment and its application,''
\newblock in {\em Dagstuhl Follow-Ups}. Schloss Dagstuhl-Leibniz-Zentrum fuer
  Informatik, 2012, vol.~3.

\bibitem{gao2020personalized}
Xiaoxue Gao, Xiaohai Tian, Yi~Zhou, Rohan~Kumar Das, and Haizhou Li,
\newblock ``Personalized singing voice generation using wavernn.,''
\newblock in {\em Odyssey}, 2020, pp. 252--258.

\bibitem{gao2019speaker}
Xiaoxue Gao, Xiaohai Tian, Rohan~Kumar Das, Yi~Zhou, and Haizhou Li,
\newblock ``Speaker-independent spectral mapping for speech-to-singing
  conversion,''
\newblock in {\em IEEE APSIPA ASC}, 2010, pp. 159--164.

\bibitem{vijayan2018analysis}
Karthika Vijayan, Xiaoxue Gao, and Haizhou Li,
\newblock ``Analysis of speech and singing signals for temporal alignment,''
\newblock in {\em IEEE APSIPA ASC}, 2018, pp. 1893--1898.

\bibitem{sharma2020automatic}
Bidisha Sharma and Ye~Wang,
\newblock ``Automatic evaluation of song intelligibility using singing adapted
  stoi and vocal-specific features,''
\newblock {\em IEEE/ACM Transactions on Audio, Speech, and Language
  Processing}, vol. 28, pp. 319--331, 2020.

\bibitem{ibrahim2017intelligibility}
Karim~M Ibrahim, David Grunberg, Kat Agres, Chitralekha Gupta, and Ye~Wang,
\newblock ``Intelligibility of sung lyrics: A pilot study.,''
\newblock in {\em ISMIR}, 2017, pp. 686--693.

\bibitem{stoller2019}
D.~Stoller, S.~Durand, and S.~Ewert,
\newblock ``End-to-end lyrics alignment for polyphonic music using an
  audio-to-character recognition model,''
\newblock in {\em IEEE ICASSP}, 2019, pp. 181--185.

\bibitem{9054567}
C.~{Gupta}, E.~{Yılmaz}, and H.~{Li},
\newblock ``Automatic lyrics alignment and transcription in polyphonic music:
  Does background music help?,''
\newblock in {\em IEEE ICASSP}, 2020, pp. 496--500.

\bibitem{gupta2019}
C.\* Gupta, B.\* Sharma, H.~Li, and Y.~Wang,
\newblock ``Automatic lyrics-to-audio alignment on polyphonic music using
  singing-adapted acoustic models,''
\newblock in {\em IEEE ICASSP}, 2019, pp. 396--400.

\bibitem{mesaros2010automatic}
A.~Mesaros and T.~Virtanen,
\newblock ``Automatic recognition of lyrics in singing,''
\newblock {\em EURASIP Journal on Audio, Speech, and Music Processing}, vol.
  2010, no. 1, pp. 546047, 2010.

\bibitem{dzhambazov2015modeling}
G.~B. Dzhambazov and X.~Serra,
\newblock ``Modeling of phoneme durations for alignment between polyphonic
  audio and lyrics,''
\newblock in {\em Sound and Music Computing Conference}, 2015, pp. 281--286.

\bibitem{fujihara2011lyricsynchronizer}
H.~Fujihara, M.~Goto, J.~Ogata, and H.~G. Okuno,
\newblock ``Lyricsynchronizer: Automatic synchronization system between musical
  audio signals and lyrics,''
\newblock {\em IEEE Journal of Selected Topics in Signal Processing}, vol. 5,
  no. 6, pp. 1252--1261, 2011.

\bibitem{gao2022genre}
Xiaoxue Gao, Chitralekha Gupta, and Haizhou Li,
\newblock ``Genre-conditioned acoustic models for automatic lyrics
  transcription of polyphonic music,''
\newblock in {\em IEEE ICASSP}, 2022, pp. 791--795.

\bibitem{demirel2021low}
Emir Demirel, Sven Ahlb{\"a}ck, and Simon Dixon,
\newblock ``Low resource audio-to-lyrics alignment from polyphonic music
  recordings,''
\newblock in {\em IEEE ICASSP}, 2021, pp. 586--590.

\bibitem{demirel2020automatic}
Emir Demirel, Sven Ahlb{\"a}ck, and Simon Dixon,
\newblock ``Automatic lyrics transcription using dilated convolutional neural
  networks with self-attention,''
\newblock in {\em IEEE IJCNN}, 2020, pp. 1--8.

\bibitem{gupta2022overview}
Chitralekha Gupta, Haizhou Li, and Masataka Goto,
\newblock ``Deep learning approaches in topics of singing information
  processing,''
\newblock {\em IEEE/ACM Transactions on Audio, Speech, and Language
  Processing}, vol. 30, pp. 2422--2451, 2022.

\bibitem{stoller2018wave}
Daniel Stoller, Sebastian Ewert, and Simon Dixon,
\newblock ``Wave-u-net: A multi-scale neural network for end-to-end audio
  source separation,''
\newblock in {\em ISMIR}, 2018, pp. 334--340.

\bibitem{lluis2018end}
Francesc Llu{\'\i}s, Jordi Pons, and Xavier Serra,
\newblock ``End-to-end music source separation: Is it possible in the waveform
  domain?,''
\newblock in {\em Interspeech}, 2019, pp. 4619--4623.

\bibitem{samuel2020meta}
David Samuel, Aditya Ganeshan, and Jason Naradowsky,
\newblock ``Meta-learning extractors for music source separation,''
\newblock in {\em IEEE ICASSP}, 2020, pp. 816--820.

\bibitem{defossez2019music}
Alexandre D{\'e}fossez, Nicolas Usunier, L{\'e}on Bottou, and Francis Bach,
\newblock ``Music source separation in the waveform domain,''
\newblock {\em arXiv preprint arXiv:1911.13254}, 2019.

\bibitem{stoter19}
Fabian-Robert St{\"o}ter, Stefan Uhlich, Antoine Liutkus, and Yuki Mitsufuji,
\newblock ``Open-unmix-a reference implementation for music source
  separation,''
\newblock {\em Journal of Open Source Software}, vol. 4, no. 41, pp. 1667,
  2019.

\bibitem{takahashi2018mmdenselstm}
Naoya Takahashi, Nabarun Goswami, and Yuki Mitsufuji,
\newblock ``Mmdenselstm: An efficient combination of convolutional and
  recurrent neural networks for audio source separation,''
\newblock in {\em 16th International Workshop on Acoustic Signal Enhancement}.
  IEEE, 2018, pp. 106--110.

\bibitem{stoter20182018}
Fabian-Robert St{\"o}ter, Antoine Liutkus, and Nobutaka Ito,
\newblock ``The 2018 signal separation evaluation campaign,''
\newblock in {\em International Conference on Latent Variable Analysis and
  Signal Separation}. Springer, 2018, pp. 293--305.

\bibitem{MUSDB18}
Zafar Rafii, Antoine Liutkus, Fabian-Robert St{\"o}ter, Stylianos~Ioannis
  Mimilakis, and Rachel Bittner,
\newblock ``The {MUSDB18} corpus for music separation,''
\newblock Access on: Dec, 2017. [ONLINE] Available:
  https://doi.org/10.5281/zenodo.1117372. DOI: 10.5281/zenodo.1117372.

\bibitem{hennequin2020spleeter}
Romain Hennequin, Anis Khlif, Felix Voituret, and Manuel Moussallam,
\newblock ``Spleeter: a fast and efficient music source separation tool with
  pre-trained models,''
\newblock {\em Journal of Open Source Software}, vol. 5, no. 50, pp. 2154,
  2020.

\bibitem{takahashi2020d3net}
Naoya Takahashi and Yuki Mitsufuji,
\newblock ``D3net: Densely connected multidilated densenet for music source
  separation,''
\newblock {\em arXiv preprint arXiv:2010.01733}, 2020.

\bibitem{kong2021decoupling}
Qiuqiang Kong, Yin Cao, Haohe Liu, Keunwoo Choi, and Yuxuan Wang,
\newblock ``Decoupling magnitude and phase estimation with deep resunet for
  music source separation,''
\newblock in {\em ISMIR}, 2021, pp. 342--349.

\bibitem{pretet2019singing}
Laure Pr{\'e}tet, Romain Hennequin, Jimena Royo-Letelier, and Andrea Vaglio,
\newblock ``Singing voice separation: A study on training data,''
\newblock in {\em IEEE ICASSP}, 2019, pp. 506--510.

\bibitem{gao2022self}
Xiaoxue Gao, Xianghu Yue, and Haizhou Li,
\newblock ``Self-transriber: Few-shot lyrics transcription with
  self-training,''
\newblock {\em arXiv preprint arXiv:2211.10152}, 2022.

\bibitem{povey2011kaldi}
A.~Povey, D.and~Ghoshal, G.~Boulianne, L.~Burget, O.~Glembek, N.~Goel,
  M.~Hannemann, P.~Motlicek, Y.~Qian, P.~Schwarz, J.~Silovsky, G.~Stemmer, and
  K.~Vesely,
\newblock ``The {Kaldi} speech recognition toolkit,''
\newblock in {\em ASRU}, 2011.

\bibitem{dabike2019automatic}
Gerardo~Roa Dabike and Jon Barker,
\newblock ``Automatic lyric transcription from karaoke vocal tracks: Resources
  and a baseline system.,''
\newblock in {\em Interspeech}, 2019, pp. 579--583.

\bibitem{gao2021tran}
Xiaoxue Gao, Chitralekha Gupta, and Haizhou Li,
\newblock ``Automatic lyrics transcription of polyphonic music with
  lyrics-chord multi-task learning,''
\newblock {\em IEEE/ACM Transactions on Audio, Speech, and Language
  Processing}, vol. 30, pp. 2280--2294, 2022.

\bibitem{mesaros2009adaptation}
Annamaria Mesaros and Tuomas Virtanen,
\newblock ``Adaptation of a speech recognizer for singing voice,''
\newblock in {\em 2009 17th European Signal Processing Conference}. IEEE, 2009,
  pp. 1779--1783.

\bibitem{mesaros2013singing}
Annamaria Mesaros,
\newblock ``Singing voice identification and lyrics transcription for music
  information retrieval invited paper,''
\newblock in {\em IEEE Conference on Speech Technology and Human-Computer
  Dialogue}, 2013, pp. 1--10.

\bibitem{kawai2016speech}
Dairoku Kawai, Kazumasa Yamamoto, and Seiichi Nakagawa,
\newblock ``Speech analysis of sung-speech and lyric recognition in monophonic
  singing,''
\newblock in {\em IEEE ICASSP}, 2016, pp. 271--275.

\bibitem{kruspe2016bootstrapping}
A.~M. Kruspe,
\newblock ``Bootstrapping a system for phoneme recognition and keyword spotting
  in unaccompanied singing.,''
\newblock in {\em ISMIR}, 2016, pp. 358--364.

\bibitem{gupta2018automatic}
Chitralekha Gupta, Haizhou Li, and Ye~Wang,
\newblock ``Automatic pronunciation evaluation of singing.,''
\newblock in {\em Interspeech}, 2018, pp. 1507--1511.

\bibitem{tsai2018transcribing}
Che-Ping Tsai, Yi-Lin Tuan, and Lin-shan Lee,
\newblock ``Transcribing lyrics from commercial song audio: the first step
  towards singing content processing,''
\newblock in {\em IEEE ICASSP}, 2018, pp. 5749--5753.

\bibitem{gupta2018semi}
Chitralekha Gupta, Rong Tong, Haizhou Li, and Ye~Wang,
\newblock ``Semi-supervised lyrics and solo-singing alignment.,''
\newblock in {\em ISMIR}, 2018, pp. 600--607.

\bibitem{dabikesheffield}
Gerardo~Roa Dabike and Jon Barker,
\newblock ``The sheffield university system for the {MIREX} 2020: Lyrics
  transcription task,''
\newblock in {\em MIREX}, 2020.

\bibitem{guptalyricsAdapt}
Chitralekha Gupta, Bidisha Sharma, Haizhou Li, and Ye~Wang,
\newblock ``Lyrics-to-audio alignment using singing-adapted acoustic models and
  non-vocal suppression,''
\newblock {\em Music Inf. Retrieval Eval. eXchange Audio-Lyrics Alignment
  Challenge}, 2022,
\newblock Accessed on: Jul. 19, 2022. [Online]. Available: https://www.
  music-ir.org/mirex/abstracts/2018/GSLW1.pdf.

\bibitem{gupta2019acoustic}
C.~Gupta, E.~Y{\i}lmaz, and H.~Li,
\newblock ``Acoustic modeling for automatic lyrics-to-audio alignment,''
\newblock in {\em Interspeech}, 2019.

\bibitem{ahlback2021mstre}
Emir Demirel, Sven Ahlb{\"a}ck, and Simon Dixon,
\newblock ``Mstre-net: Multistreaming acoustic modeling for automatic lyrics
  transcription,''
\newblock in {\em ISMIR}, 2021, pp. 151--158.

\bibitem{gao2022music}
Xiaoxue Gao, Chitralekha Gupta, and Haizhou Li,
\newblock ``Music-robust automatic lyrics transcription of polyphonic music,''
\newblock in {\em Sound and Music Computing}, 2022, pp. 325--332.

\bibitem{graves2014towards}
Alex Graves and Navdeep Jaitly,
\newblock ``Towards end-to-end speech recognition with recurrent neural
  networks,''
\newblock in {\em International conference on machine learning}. PMLR, 2014,
  pp. 1764--1772.

\bibitem{miao2015eesen}
Yajie Miao, Mohammad Gowayyed, and Florian Metze,
\newblock ``Eesen: End-to-end speech recognition using deep rnn models and
  wfst-based decoding,''
\newblock in {\em IEEE ASRU}, 2015, pp. 167--174.

\bibitem{chorowski2014end}
Jan Chorowski, Dzmitry Bahdanau, Kyunghyun Cho, and Yoshua Bengio,
\newblock ``End-to-end continuous speech recognition using attention-based
  recurrent nn: First results,''
\newblock in {\em NIPS Workshop on Deep Learning, December}, 2014.

\bibitem{chan2016listen}
William Chan, Navdeep Jaitly, Quoc Le, and Oriol Vinyals,
\newblock ``Listen, attend and spell: A neural network for large vocabulary
  conversational speech recognition,''
\newblock in {\em IEEE ICASSP}, 2016, pp. 4960--4964.

\bibitem{kim2017joint}
Suyoun Kim, Takaaki Hori, and Shinji Watanabe,
\newblock ``Joint ctc-attention based end-to-end speech recognition using
  multi-task learning,''
\newblock in {\em IEEE ICASSP}, 2017, pp. 4835--4839.

\bibitem{nakatani2019improving}
Shigeki Karita, Nelson Enrique~Yalta Soplin, Shinji Watanabe, Marc Delcroix,
  Atsunori Ogawa, and Tomohiro Nakatani,
\newblock ``Improving transformer-based end-to-end speech recognition with
  connectionist temporal classification and language model integration,''
\newblock in {\em Interspeech}, 2019, pp. 1408--1412.

\bibitem{jansson2017singing}
A.~Jansson, E.~J. Humphrey, N.~Montecchio, R.~M. Bittner, A.~Kumar, and
  T.~Weyde,
\newblock ``Singing voice separation with deep {U-Net} convolutional
  networks,''
\newblock in {\em ISMIR}, 2017, pp. 745--751.

\bibitem{chandna2017monoaural}
P.~Chandna, M.~Miron, J.~Janer, and E.~G{\'o}mez,
\newblock ``Monoaural audio source separation using deep convolutional neural
  networks,''
\newblock in {\em International Conference on Latent Variable Analysis and
  Signal Separation}. Springer, 2017, pp. 258--266.

\bibitem{stoter2019open}
Fabian-Robert St{\"o}ter, Stefan Uhlich, Antoine Liutkus, and Yuki Mitsufuji,
\newblock ``Open-unmix-a reference implementation for music source
  separation,''
\newblock {\em Journal of Open Source Software}, vol. 4, no. 41, pp. 1667,
  2019.

\bibitem{choi2018phase}
Hyeong-Seok Choi, Jang-Hyun Kim, Jaesung Huh, Adrian Kim, Jung-Woo Ha, and
  Kyogu Lee,
\newblock ``Phase-aware speech enhancement with deep complex {U}-{N}et,''
\newblock in {\em International Conference on Learning Representations}, 2018,
\newblock https://openreview.net/forum?id=SkeRTsAcYm.

\bibitem{wang2018supervised}
DeLiang Wang and Jitong Chen,
\newblock ``Supervised speech separation based on deep learning: An overview,''
\newblock {\em IEEE/ACM Transactions on Audio, Speech, and Language
  Processing}, vol. 26, no. 10, pp. 1702--1726, 2018.

\bibitem{tan2019complex}
Ke~Tan and DeLiang Wang,
\newblock ``Complex spectral mapping with a convolutional recurrent network for
  monaural speech enhancement,''
\newblock in {\em IEEE ICASSP}, 2019, pp. 6865--6869.

\bibitem{mitsufuji2021music}
Yuki Mitsufuji, Giorgio Fabbro, Stefan Uhlich, Fabian-Robert St{\"o}ter,
  Alexandre D{\'e}fossez, Minseok Kim, Woosung Choi, Chin-Yun Yu, and Kin-Wai
  Cheuk,
\newblock ``Music demixing challenge 2021,''
\newblock {\em Frontiers in Signal Processing}, vol. 1, pp. 18, 2022.

\bibitem{vaswani2017attention}
Ashish Vaswani, Noam Shazeer, Niki Parmar, Jakob Uszkoreit, Llion Jones,
  Aidan~N. Gomez, Lukasz Kaiser, and Illia Polosukhin,
\newblock ``Attention is all you need,''
\newblock in {\em Neural Information Processing Systems}, 2017, pp. 5998--6008.

\bibitem{he2016deep}
Kaiming He, Xiangyu Zhang, Shaoqing Ren, and Jian Sun,
\newblock ``Deep residual learning for image recognition,''
\newblock in {\em IEEE conference on computer vision and pattern recognition},
  2016, pp. 770--778.

\bibitem{ba2016layer}
Jimmy~Lei Ba, Jamie~Ryan Kiros, and Geoffrey~E Hinton,
\newblock ``Layer normalization,''
\newblock {\em arXiv preprint arXiv:1607.06450}, 2016.

\bibitem{meseguer2018dali}
G.~Meseguer-Brocal, A.~Cohen-Hadria, and G.~Peeters,
\newblock ``Dali: A large dataset of synchronized audio, lyrics and notes,
  automatically created using teacher-student machine learning paradigm,''
\newblock in {\em ISMIR}, 2018, pp. 431--437.

\bibitem{hansen2012recognition}
J.~K. Hansen,
\newblock ``Recognition of phonemes in a-cappella recordings using temporal
  patterns and mel frequency cepstral coefficients,''
\newblock in {\em Sound and Music Computing}, 2012, pp. 494--499.

\bibitem{mauch2010lyrics}
M.~Mauch, H.~Fujihara, and M.~Goto,
\newblock ``Lyrics-to-audio alignment and phrase-level segmentation using
  incomplete internet-style chord annotations,''
\newblock in {\em Sound and Music Computing}, 2010, pp. 9--16.

\bibitem{watanabe2018espnet}
S.~Watanabe, T.~Hori, S.~Karita, T.~Hayashi, J.~Nishitoba, Y.~Unno, N.~Enrique
  Yalta~Soplin, J.~Heymann, M.~Wiesner, N.~Chen, A.~Renduchintala, and
  T.~Ochiai,
\newblock ``Espnet: End-to-end speech processing toolkit,''
\newblock in {\em Interspeech}, 2018, pp. 2207--2211.

\bibitem{hori2018end}
Takaaki Hori, Jaejin Cho, and Shinji Watanabe,
\newblock ``End-to-end speech recognition with word-based rnn language
  models,''
\newblock in {\em IEEE SLT}, 2018, pp. 389--396.

\bibitem{panayotov2015librispeech}
V.~Panayotov, G.~Chen, D.~Povey, and S.~Khudanpur,
\newblock ``Librispeech: an {ASR} corpus based on public domain audio books,''
\newblock in {\em ICASSP}, 2015, pp. 5206--5210.

\bibitem{wang2016joint}
Zhong-Qiu Wang and DeLiang Wang,
\newblock ``A joint training framework for robust automatic speech
  recognition,''
\newblock {\em IEEE/ACM Transactions on Audio, Speech, and Language
  Processing}, vol. 24, no. 4, pp. 796--806, 2016.

\bibitem{chang2019mimo}
Xuankai Chang, Wangyou Zhang, Yanmin Qian, Jonathan Le~Roux, and Shinji
  Watanabe,
\newblock ``Mimo-speech: End-to-end multi-channel multi-speaker speech
  recognition,''
\newblock in {\em IEEE ASRU}, 2019, pp. 237--244.

\bibitem{wu2021investigation}
Jian Wu, Zhuo Chen, Sanyuan Chen, Yu~Wu, Takuya Yoshioka, Naoyuki Kanda, Shujie
  Liu, and Jinyu Li,
\newblock ``Investigation of practical aspects of single channel speech
  separation for asr,''
\newblock in {\em Interspeech}, 2021, pp. 3066--3070.

\bibitem{condit2015catching}
N.~Condit-Schultz and D.~Huron,
\newblock ``Catching the lyrics: intelligibility in twelve song genres,''
\newblock {\em Music Perception: An Interdisciplinary Journal}, vol. 32, no. 5,
  pp. 470--483, 2015.

\bibitem{demirel2020recursive}
Emir Demirel, Sven Ahlback, and Simon Dixon,
\newblock ``A recursive search method for lyrics alignment,''
\newblock in {\em MIREX}, 2020.

\bibitem{gaolyrics}
Xiaoxue Gao, Chitralekha Gupta, and Haizhou Li,
\newblock ``Lyrics transcription and lyrics-to-audio alignment with
  music-informed acoustic models,''
\newblock in {\em MIREX}, 2021.

\end{thebibliography}

 

\begin{IEEEbiography}[{\includegraphics[width=1in,height=1.25in,clip,keepaspectratio]{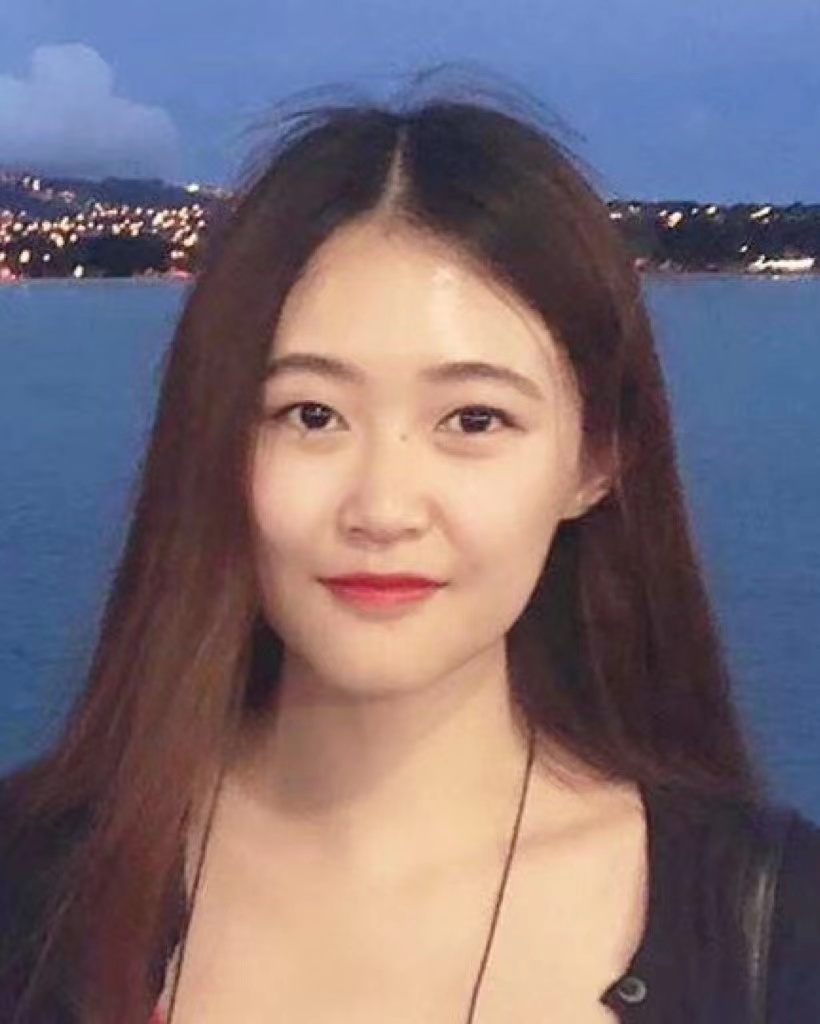}}]{Xiaoxue Gao}
 (Member, IEEE) received the Ph.D.~degree from the National University of Singapore (NUS) in 2022 and received the B.Eng. degree in Electronic Information Science and Technology from Nanjing University, China, in 2017. She is currently a post-doctoral research fellow at NUS. Her research interests include automatic lyrics transcription, speech recognition, speech-to-singing conversion, singing information processing, music information retrieval and multi-modal processing.
\end{IEEEbiography}

\vspace{11pt}

\begin{IEEEbiography}[{\includegraphics[width=1in,height=1.25in,clip,keepaspectratio]{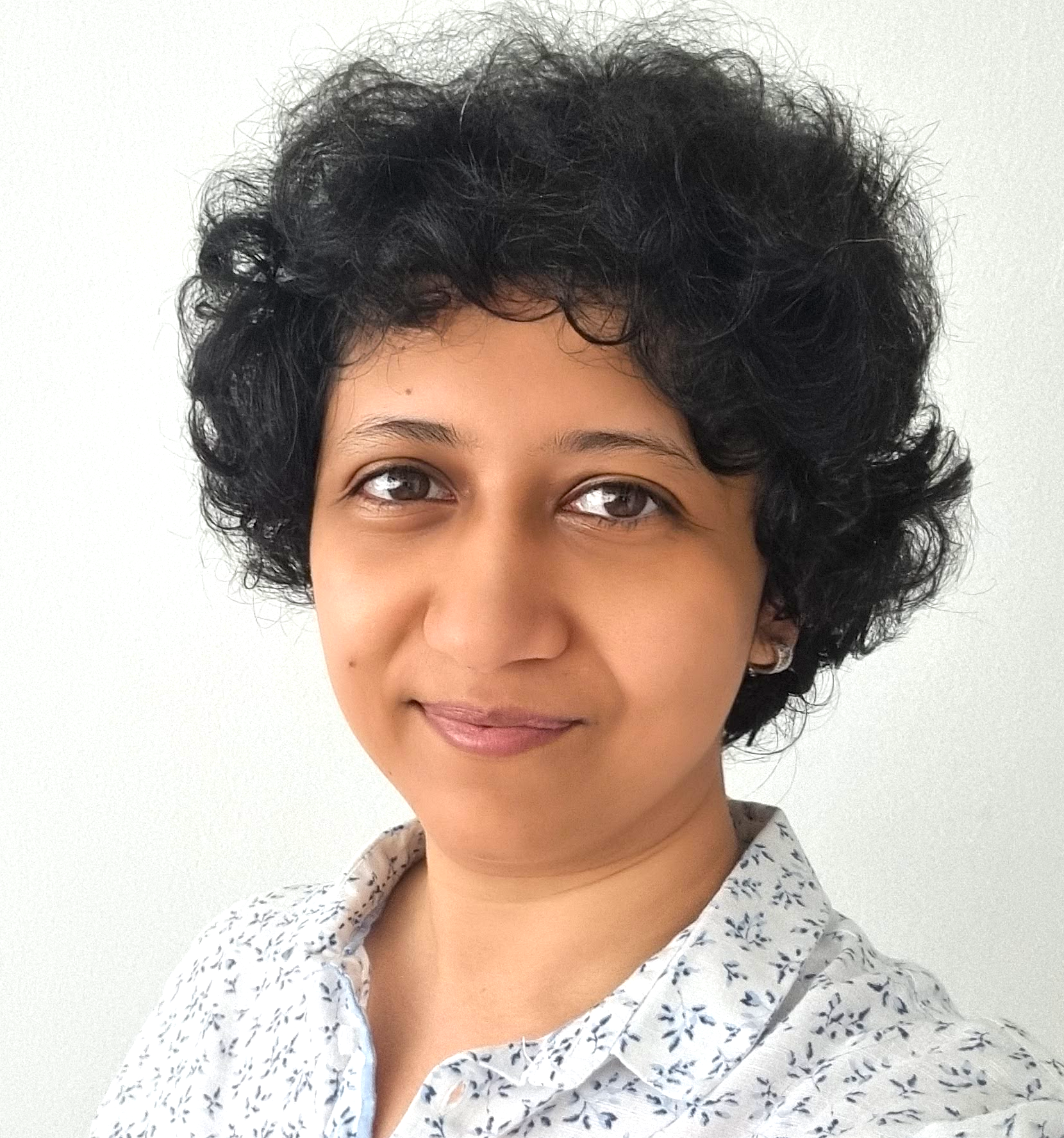}}]{Chitralekha Gupta} (Member, IEEE) received the Ph.D.~degree from the National University of Singapore (NUS) in 2019, Masters in engineering degree from Indian Institute of Technology Bombay (IIT-B), India in 2011, and Bachelors in engineering degree from Maharaja Sayajirao University (MSU) Baroda, India in 2008. She is currently a Senior Research Fellow at the School of Computing, NUS. She has been awarded the start-up grant of Graduate Research Innovation Program NUS and has founded the music tech company MuSigPro Pte. Ltd. in Singapore in 2019. She is the recipient of the NUS Dean's Graduate Research Achievement Award 2018, the NUS School of Computing Innovation Prize 2018, and the Best Student Paper Award in APSIPA 2017. Her current research interests are neural audio synthesis, affective modeling of audio, singing voice analysis, and applications of ASR in music. She has played an active role in the organizing committees of several international conferences and challenges including ISMIR 2022, ICASSP 2022, and MIREX 2020.
\end{IEEEbiography}

\vspace{11pt}

\begin{IEEEbiography}[{\includegraphics[width=1in,height=1.25in,clip,keepaspectratio]{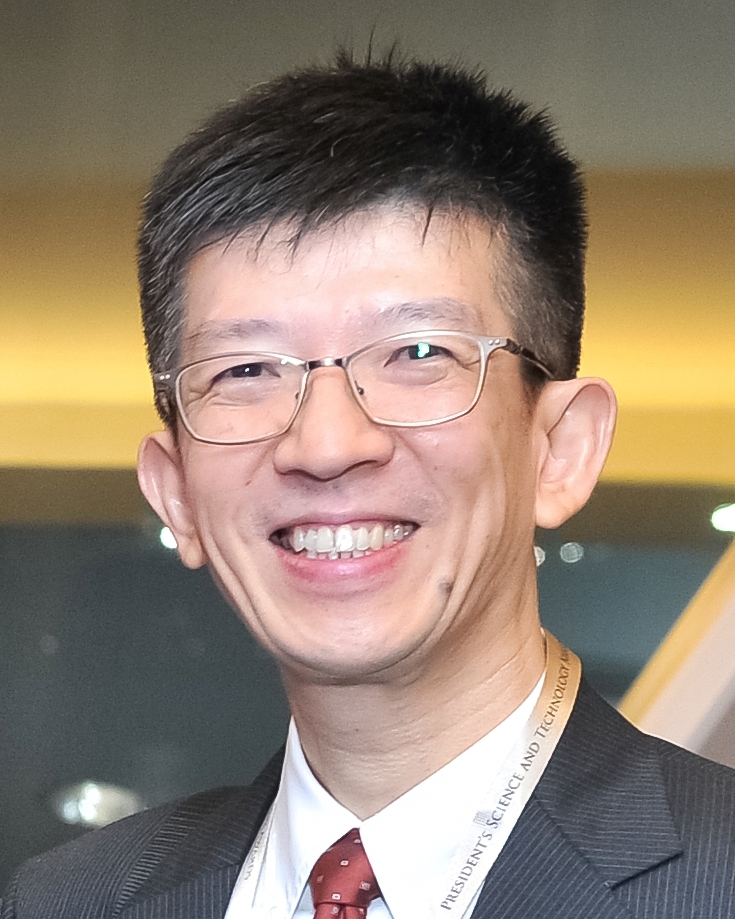}}]{Haizhou Li}
 (Fellow, IEEE) received the B.Sc., M.Sc., and Ph.D. degrees in electrical and electronic engineering from the South China University of Technology, Guangzhou, China, in 1984, 1987, and 1990 respectively. He is currently a Presidential Chair Professor and the Executive Dean of the School of Data Science, The Chinese University of Hong Kong, Shenzhen, China. He is also an Adjunct Professor with the Department of Electrical and Computer Engineering, National University of Singapore, Singapore. Prior to that, he taught with the University of Hong Kong, Hong Kong, (1988–1990) and South China University of Technology, (1990–1994). He was a Visiting Professor with CRIN in France (1994–1995), Research Manager with the AppleISS Research Centre (1996–1998), Research Director with Lernout \& Hauspie Asia Pacific (1999–2001), a Vice President with InfoTalk Corp. Ltd. (2001–2003), and the Principal Scientist and Department Head of Human Language Technology with the Institute for Infocomm Research, Singapore (2003–2016). His research interests include automatic speech recognition, speaker and language recognition, natural language processing. Dr. Li was an Editor-in-Chief of IEEE/ACM Transactions on Audio, Speech and Language Processing (2015–2018), a Member of the Editorial Board of Computer Speech and Language since 2012, an elected Member of IEEE Speech and Language Processing Technical Committee (2013–2015), the President of the International Speech Communication Association (2015–2017), the President of Asia Pacific Signal and Information Processing Association (2015–2016), and the President of Asian Federation of Natural Language Processing (2017–2018). He was the General Chair of ACL 2012, INTERSPEECH 2014, ASRU 2019 and ICASSP 2022. Dr. Li is a Fellow of the ISCA, and a Fellow of the Academy of Engineering Singapore. He was the recipient of the National Infocomm Award 2002, and the President's Technology Award 2013 in Singapore. He was named one of the two Nokia Visiting Professors in 2009 by the Nokia Foundation, and U Bremen Excellence Chair Professor in 2019.
\end{IEEEbiography}

\vfill
\end{document}